\documentclass[11pt,a4paper]{article}
\usepackage{jheppub}
\usepackage[utf8]{inputenc}
\usepackage[T1]{fontenc}
\usepackage{hyperref}
\usepackage{amsfonts}
\usepackage{amsmath}
\usepackage{graphicx}
\usepackage{mathrsfs}
\usepackage{color}
\usepackage{tabu}
\usepackage{mathtools}
\usepackage{url}
\usepackage[titletoc,title]{appendix}
\usepackage{soul}
\usepackage{multicol}

\newcommand{\scalar}[2]{\left\langle #1, \, #2 \right\rangle}

\def\doubleunderline#1{\underline{\underline{#1}}}
\title{AdS instability: resonant system for gravitational perturbations of AdS${}_5$ in the cohomogeneity-two biaxial Bianchi IX ansatz}
\author{Dominika Hunik-Kostyra,}
\author{Andrzej Rostworowski}
\affiliation{Institute of Theoretical Physics, Jagiellonian University, Łojasiewicza 11, 30-348 Krakow, Poland}
\emailAdd{hunik.dominika@gmail.com}
\emailAdd{arostwor@th.if.uj.edu.pl}
\date{\today}

\abstract{
We consider five-dimensional, vacuum Einstein equations with negative cosmological constant within cohomogenity-two biaxial Bianchi IX ansatz. This model allows to investigate the stability of AdS without adding any matter to the energy-momentum tensor, thus analyzing instability of genuine gravtational degrees of freedom. We derive the resonant system and identify vanishing secular terms. The results resemble those obtained for Einstein equations coupled to a spherically-symmetric, massless scalar field, backing the evidence that the scalar field model captures well the relevant features of AdS instability problem. We also list recurrence relations for the interaction coefficients of the resonant system, which might be useful in both numerical simulations and further analytical studies.
} 

\begin{document}
\maketitle

\section{Introduction}
\label{Introduction}

Over the past two decades asymptotically anti-de Sitter (aAdS)
spacetimes have received a great deal of attention, primarily due to
the AdS/CFT correspondence which is the conjectured duality between
aAdS spacetimes and conformal field theories.  The distinctive feature
of aAdS spacetimes, on which the very concept of duality rests, is a
time-like conformal boundary at spatial and null infinity, where it is
necessary to specify boundary conditions in order to define the
deterministic evolution. For energy conserving boundary conditions the
conformal boundary acts as a mirror at which massless waves
propagating outwards bounce off and return to the bulk.  Therefore,
the key mechanism stabilizing the evolution of asymptotically flat
spacetimes -- dispersion of energy by radiation -- is absent in aAdS
spacetimes. For this reason the problem of nonlinear stability of the
pure AdS spacetime (which is the ground state among aAdS spacetimes)
is particularly challenging.  The first conjecture, based on numerical
evidence and heuristic arguments, about AdS being unstable against
gravitational collapse (black hole formation) under arbitrarily small
perturbations came from Bizo\'n and one of the present authors
(2011)~\cite{br_PRL107}. 
More precisely, in a toy model of the spherically symmetric massless
scalar field minimally coupled to gravity with a negative cosmological
constant in four \cite{br_PRL107} and higher dimensions \cite{jrb_PRD84} the
numerical simulations showed that there is a class of
arbitrarily small perturbations of AdS that evolve into a black hole
on the time-scale $\mathcal{O}(\varepsilon^{-2})$, where $\varepsilon$
measures the amplitude of the perturbation. Moreover, on the basis of nonlinear
perturbation analysis it was argued that this instability is due to a
resonant transfer of energy from low to high frequencies, or
equivalently, from coarse to fine spatial scales~
\footnote{
Perturbation can be decomposed at any instant of time as an infinite sum of (a complete set of) linear AdS eigen modes with time dependent coefficients. By the resonant transfer of energy we mean that the conserved energy of the system leaks to the modes with arbitrarily high frequencies, even if initially distributed among low frequency modes. 
},  
until eventually an apparent horizon forms~
\footnote{Remarkably, a proof of AdS instability for a model Einstein--null dust system has recently been given \cite{m_1704.08681, m_1812.04268}. The proof does not set the time-scale of a black hole formation, in particular it does not relate this time-scale to the amplitude of initial perturbation. The position space analysis, similar in spirit to that of the proof \cite{m_1704.08681, m_1812.04268}, was attempted for the first time in the context of AdS--Einstein--massless scalar field model in \cite{dfly_JHEP1508}.}. 

Further studies of this and similar models confirmed (see \cite{dkfk_PRL114, df_JHEP1512} for independent, reliable long-time numerical integration of Einstein equations in Einstein--scalar fields models) and extended the
findings of \cite{br_PRL107, jrb_PRD84} providing important new insights concerning the coexistence of unstable (turbulent) and stable (quasiperiodic) regimes
of evolution~
\footnote{It is quite remarkable that even if AdS solution itself is not stable there exist globally regular, aAdS solutions of Einstein equations that, as numerical evidence shows, are immune to the instability discovered in \cite{br_PRL107} at least on $\mathcal{O}\left( \varepsilon^{-k}\right)$ time-scale. These are time-periodic solutions in Einstein-scalar fields models \cite{mr_PRL111, m_PhD, bll_PRD87, ffg_PRD92}, in the presently studied cohomogenity-two biaxial Bianchi IX ansatz \cite{m_PhD}, and time-periodic (in axial symmetry) or helically symmetric (outside axial symmetry) globally regular aAdS vacuum solutions (geons) \cite{dhs_CQG29, hs, mfgf_CQG34, r_PRD95, ff_PRD96}. The stability of the laters is assumed on the ground of numerical evidence for the stability of  the formers.} 
(see \cite{ce_FP64} for a brief review and references).

Still, there are two major downsides of all reported evidence for AdS instability, based on numerical integration of Einstein
equations.  First, the arguments of \cite{br_PRL107,jrb_PRD84} and
following works were based on extrapolation of the observed scaling
$\mathcal{O}(\varepsilon^{-2})$ in time of resonant energy transfers
between the modes and ultimately collapse times for finite small
values of $\varepsilon$, cf. Fig.~2 in \cite{br_PRL107}, but the limit
$\varepsilon \rightarrow 0$, with the instability time scale
$\varepsilon^{-2}$, is obviously inaccessible to
numerical simulation.  Second, the numerical integration of Einstein
equations on the time scales long enough to provide convincing
evidence for the AdS instability seems tractable only under some
simplifying symmetry assumptions.  Thus most numerical simulations
were restricted to spherical symmetry where adding some matter
(usually in the form of massless scalar field) was necessary to evade
Birkhoff's theorem and generate the dynamics, so that no
gravitational degrees of freedom were excited~
\footnote{It is expected on the grounds of perturbative analysis of \cite{dhs_CQG29} that the mechanism for instability of AdS in the vacuum case (pure gravity) is the same as in the model case \cite{br_PRL107}. The first steps to run simulations outside spherical symmetry in $2+1$ dimensional setting were done in \cite{bfkr_PRL119}, but the "big" perturbations (collapsing after few bounces) can not provide the evidence for the scaling $\mathcal{O}(\varepsilon^{-2})$.}. 
The first numerical evidence for AdS instability in vacuum
Einstein equations with negative cosmological constant in five
dimensions within the cohomogenity-two biaxial Bianchi IX ansatz was
reported in \cite{br_APPB48} (in fact, this model was studied in
parallel with \cite{br_PRL107}, but the results were published only recently).  Indeed, one may avoid assumptions about
spherical symmetry and still keep effectively $1+1$ dimensional
setting using the fact that Birkhoff's theorem can be evaded in five
and higher odd spacetime dimensions as was observed for the first time
in \cite{bcs_PRL95} in the context of critical collapse for
asymptotically flat spacetimes. Odd-dimensional spheres admit
non-round homogeneous metrics. Here we focus on $4+1$ dimensional aAdS spacetimes with the boundary $R \times S^3$. The key idea is to use the homogeneous
metric on $S^3$, which takes the form
\begin{equation}
g_{S^3} = e^{2B} \sigma_1^2+e^{2C} \sigma_2^2 + e^{2D}\sigma_3^2\;,
\end{equation}
as an angular part of the five-dimensional metric (cohomogenity-two triaxial Bianchi IX ansatz)~\cite{bcs_PRL95}
\begin{equation}
\label{bcs_ansatz}
ds^2= -A e^{-2\delta}dt^2 +A^{-1}dr^2 + \frac{1}{4}r^2 g_{S^3}\;.
\end{equation}
Here $\sigma_k$ are left-invariant one-forms on $SU(2)$
\begin{equation}
\sigma_1+i\sigma_2 = e^{i\psi}(\cos \theta d\phi+id\theta)\;,\;\;\sigma_3 = d\psi-\sin\theta d\phi\;
\end{equation}
and $A$, $\delta$, $B$, $C$ are functions of time and radial coordinates. 
In the biaxial case we have $B=C$.

To deal with the problem of extrapolation of results of numerical integration of Einstein equations to the $\varepsilon \rightarrow 0$ limit i.e. to track the effects of a small perturbation over large time scales Balasubramanian, Buchel, Green, Lehner \& Liebling (2014)~\cite{bbgll_PRL113} and Craps, Evnin \& Vanhoof (2014)~\cite{cev_JHEP1410} introduced new resummation schemes of a na\"ive nonlinear perturbation expansion based on multi-time framework and renormalization group methods, respectively. The Secs.~1~and~2 of \cite{cev_JHEP1410} contain a very nice summary of the problems of na\"ive time-dependent perturbation expansion and the ways to cure them. In general, if the frequencies of linear perturbations satisfy the resonant condition, i.e. the sum or difference of two linear frequencies coincide with another linear frequency   
as is the case in AdS, then in a na\"ive perturbation expansion the secular terms, i.e. the terms that grow in time, appear. For the model \cite{br_PRL107} this happens at the third order of expansion with the appearance of $~ \varepsilon^2 t$ terms and invalidates such na\"ive expansion on the $\mathcal{O}(\varepsilon^{-2})$ timescales. Craps, Evnin \& Vanhoof (2014) \cite{cev_JHEP1410} showed how to resum such terms in the form of renormalization flow equations for the first order amplitudes and phases that in a na\"ive perturbation expansions are simply constants determined by initial data. We call such flow equations the resonant system (this name comes from the another derivation of theses equations based on averaging: the effects of all non-resonant terms average to zero and only resonant terms are important for the long-time scale dynamics \cite{cev_JHEP1501}). Such resonant system offers new ways to study the AdS stability problem \cite{cev_JHEP1501, bgll_PRD91, gmll_PRD92, bmr_PRL115, dfpy_PRD94, d_PRD100} and the studies of analogue resonant systems with simple interaction coefficients became a very active area of research on its own \cite{bcehlm_CMP353, bbe_1805.03634, ep_1808.09173}. In \cite{bmr_PRL115} the convergence between
\begin{enumerate}
\item the results of numerical integration of Einstein equation, extrapolated to the $\epsilon \rightarrow 0$ limit and,
\item the results of numerical integration of the resonant system truncated at $N$ modes, extrapolated to $N \rightarrow \infty$ limit    
\end{enumerate} 
was demonstrated. Moreover, the evidence for a blowup in finite time $\tau_H$ of solutions of the resonant system, starting from the $\varepsilon$-size initial perturbations of AdS that for Einstein equations lead to gravitational collapse at $t_H \approx \varepsilon^{-2}\tau_H$ \cite{bmr_PRL115}, provided a very strong argument for the extrapolation $\varepsilon \rightarrow 0$, made in \cite{br_PRL107}, to be correct.   

In this work we construct the resonant system for the AdS-Einstein equations with cohomogenity-two biaxial Bianchi IX ansatz studied in \cite{br_APPB48}. Our motivation is two-fold. First, we want to strengthen the evidence for the AdS instability in vacuum Einstein equations, as was done in \cite{bmr_PRL115} for the model with the scalar field. Constructing the resonant system itself is the first step in this direction. Second, with the recently described systematic approach to nonlinear gravitational perturbations \cite{r_PRD95, r_PRD96, ff_PRD96, ds_CQG35} it should be possible to obtain the resonant system for arbitrary gravitational perturbation. Thus we treat the construction of the resonant system under simplifying symmetry assumptions (\ref{bcs_ansatz}) as a test-case and feasibility study for this ambitious project.

The work is organized as follows. In the section~\ref{setup} we setup our system and follow the method of Craps, Evnin \& Vanhoof (2014)~\cite{cev_JHEP1410} to obtain a resonant system for the Einstein equations; we also discuss the vanishing of two classes of secular terms, allowed by the AdS resonant spectrum but in fact not present in the resonant system, analogously to the massless scalar field case \cite{cev_JHEP1410, cev_JHEP1501}. In the section~\ref{recurrences} we derive the recurrence relations for the (interaction) coefficients in the resonant system that can be useful both to calculate their numerical values and study their asymptotic behavior. In the section~\ref{PreliminaryNumericalResults} we comment very briefly on the preliminary results of numerical integration of the resonant system. Some technical details of the calculations presented in Sec.~\ref{setup} are delegated to two appendices.

\section{Setup of the system}
\label{setup}
We consider $d+1$ dimensional vacuum Einstein equations with a negative cosmological constant
\begin{equation} 
\label{Einsteineq1}
G_{\mu\nu} +\Lambda g_{\mu\nu} = 0\;,
\end{equation}
where $\Lambda = - d(d-1)/\left(2 \ell^2 \right)$, $\ell$ is the AdS radius and $d$ stands for the number of spatial dimensions. In this work we focus on the $d=4$ case. Following \cite{bcs_PRL95}, we assume the cohomogenity-two biaxial Bianchi IX Ansatz as a gravitational perturbation of the AdS spacetime:
\begin{equation} 
\label{ansatz}
ds^2= \frac{\ell^2}{\cos^2 x} \left( -A e^{-2\delta} dt^2+A^{-1} dx^2 +\frac{1}{4} \sin^2 x (e^{2B} (\sigma_1^2 + \sigma_2^2)+e^{-4B} \sigma_3^2) \right) \; ,
\end{equation}
where $x$ is a compactified radial coordinate, $\tan x = r/\ell$, and $A$, $\delta$ and $B$ are functions of $(t,x)$ . The coordinates take the values $t\in (-\infty,\infty)$, $x\in [0, \pi/2)$. Inserting the metric~\eqref{ansatz} into~\eqref{Einsteineq1} with $\Lambda = -6/\ell^2$, we get a hyperbolic-elliptic system \cite{br_APPB48}
\begin{subequations} 
\label{Einsteineq}
\begin{align}
\label{EinsteineqB}
\dot B &= A e^{-\delta} P, 
\qquad 
\dot P = \frac{1}{\tan^3{\!x}} \left(\tan^3{\!x}\, A e^{-\delta} Q \right)'-\frac{4 e^{-\delta}}{3\sin^2{\!x}}\left(e^{-2B}-e^{-8B}\right),\\
\label{EinsteineqA}
A' &= 4 \tan{x} \, (1-A) -  2\sin{x} \cos{x} \, A \left(Q^2 + P^2 \right)
+\frac{2(4e^{-2B}-e^{-8B}-3A)}{3\tan{x}}\,,
\\
\label{Einsteineqdelta}
\delta' &= -2\sin{x} \cos{x} \left(Q^2+P^2\right) \,,
\\
\label{ad}
\dot A &= - 4 \sin{x} \cos{x} \, A^2 e^{-\delta} Q P \,,
\end{align}
\end{subequations} 
where we have introduced the auxiliary variables $Q=B'$ and $P=A^{-1} e^{\delta} \dot B$ and overdots and primes denote derivatives with respect to $t$ and $x$, respectively. The field $B$ is the only dynamical degree of freedom which plays a role similar to the spherical scalar field in \cite{br_PRL107}. If $B=0$, the only solution is the Schwarzschild-AdS family, in agreement with the Birkhoff theorem. It is convenient to define the mass function
\begin{equation}\label{mass-function}
m(t,x) = \frac{\sin^2{x}}{\cos^4{x}}\,(1-A(t,x)).
\end{equation}
From the Hamiltonian constraint (\ref{EinsteineqA}) it follows that
\begin{equation}
m'(t,x) = 2\left[A(Q^2+P^2) + \frac{1}{3\sin^2 x} \left( 3 + e^{-8B} - 4e^{-2B} \right)\right] \tan^3 x\geq 0\,.
\end{equation}
To study the problem of stability of AdS space within the ansatz (\ref{ansatz}) we need to solve the system (\ref{Einsteineq}) for small smooth initial data with finite total mass~
\footnote{Mass $M$ being finite implies $M$ being conserved as well.} 
\begin{equation}
M = \lim_{x\rightarrow\pi/2} m(t,x)
= 2 \int_0^{\pi/2} \left[A(Q^2+P^2) + \frac{1}{3\sin^2 x} \left( 3 + e^{-8B} - 4e^{-2B} \right)\right] \tan^3 x \, dx 
\end{equation}
and study the late-time behavior of its solutions. Smoothness at $x=0$ implies that
\begin{equation}\label{x=0}
   B(t,x)= b_0(t)\,x^2+\mathcal{O}(x^4), \quad\delta(t,x)= \mathcal{O}(x^4),\quad
    A(t,x)=1+\mathcal{O}(x^4),
\end{equation}
where we used normalization $\delta(t,0)=0$ to ensure  that $t$ is the proper time at the origin.
The power series  \eqref{x=0} are uniquely determined by the free function $b_0(t)$.
Smoothness at $x=\pi/2$ and finiteness of the total mass $M$ imply that  (using $\rho=x-\pi/2$)
\begin{equation}\label{pi2}
     B(t,x)= b_{\infty}(t)\, \rho^4+\mathcal{O}\left(\rho^6\right),\quad
    \delta(t,x)= \delta_{\infty}(t)+\mathcal{O}\left(\rho^8\right),\quad
    A(t,x)= 1-M \rho^4+\mathcal{O}\left(\rho^6\right)\,,
\end{equation}
where the free functions $b_{\infty}(t)$, $\delta_{\infty}(t)$, and  mass $M$ uniquely determine the power series.
  It follows from \eqref{pi2} that the asymptotic behaviour of fields at  infinity is completely fixed by the  assumptions of smoothness and finiteness of total mass, hence there is no freedom of imposing the boundary data. For the future convenience, following the conventions of~\cite{cev_JHEP1410}, we define
\begin{equation}
\mu(x) = \tan^3 x \quad \mbox{and} \quad \nu(x) = \frac{3}{\mu'(x)} = \frac{\cos^4 x}{\sin^2 x}\; .
\end{equation}

The pure AdS spacetime corresponds to $B=0,A=1,\delta=0$. Linearizing around this solution, we obtain
  \begin{equation}\label{L}
     \ddot B +L B=0 ,\qquad
    L=-\frac{1}{\mu(x)}\, \partial_x \left(\mu(x) \,\partial_x\right)+\frac{8}{\sin^2{\!x}}\,.
\end{equation}
This equation is the $\ell=2$ gravitational tensor case of the  master equation describing the evolution of linearized perturbations of AdS spacetime, analyzed in detail by Ishibashi and Wald \cite{iw_CQG21}.
The Sturm-Liouville operator $L$ is essentially self-adjoint with respect to the inner product $\scalar{f}{g}=\int_0^{\pi/2} f(x) g(x) \mu(x) \, dx$. The eigenvalues and  associated orthonormal eigenfunctions of $L$ are 
\begin{equation}
\label{eigenEq}
L \, e_k(x) = \omega_k^2 \, e_k(x), \qquad k=0,1,\dots
\end{equation} 
with
\begin{equation}\label{modes}
\omega^2_k=(6+2k)^2,\qquad e_k(x)= 2 \sqrt{\frac{(k+3)(k+4)(k+5)}{(k+1)(k+2)}}\, \sin^2{\!x} \cos^4{\!x} \,P_k^{(3,2)}(\cos{2x})\,,
\end{equation}
where $P_k^{(a,b)}(x)$ is a Jacobi polynomial of order $k$.

The eigenfunctions $e_k(x)$ fulfill the regularity conditions \eqref{x=0} and \eqref{pi2}, hence any smooth solution can be expressed as
\begin{equation}
B(t,x)=\sum\limits_{k\geq 0} b_k(t) e_k(x)\,.
\end{equation}
To quantify the transfer of energy between the modes one can introduce the linearized energy
\begin{equation}
\label{E}
E = \int_0^{\pi/2} \left( \dot B^2 + B'^2 + \frac{8}{\sin^2 x} B^2 \right) \mu(x)\, dx=\sum\limits_{k\geq 0} E_k,
\end{equation}
where $E_k=\dot b_k^2 + \omega_k^2 b_k^2$ is the linearized energy of the $k$-th mode.

\section{Construction of the resonant system}
We will look for approximate solutions of the system (\ref{Einsteineq}) with initial conditions $B(0,x) = \varepsilon f(x)$ and $\dot B(0,x) = \varepsilon g(x)$. Assuming $\varepsilon$ to be "small" we expand the metric functions $B$, $A$ and $\delta$ as series in the amplitude of the initial data:
\begin{subequations}
\label{series}
\begin{equation} \label{seriesB}
B(t,x) = \sum_{k=1}^{\infty}\varepsilon^k B_k(t,x)\;,
\end{equation}
\begin{equation} \label{seriesA}
A(t,x) = 1 + \sum_{k=2}^{\infty}\varepsilon^{k} A_{k}(t,x)\;,
\end{equation}
\begin{equation} \label{seriesdelta}
\delta(t,x) = \sum_{k=2}^{\infty}\varepsilon^{k} \delta_{k}(t,x)\;.
\end{equation}
\end{subequations}
To satisfy the initial data we take $B_1(0,x) = f(x)$, $\dot B_1(0,x) = g(x)$ and $B_k(0,x) \equiv 0$ for $k>1$. 

\subsection{First order perturbations} 

At the first order of the $\varepsilon$-expansion, the equations~(\ref{EinsteineqA},\ref{Einsteineqdelta}) are identically satisfied and the equation~(\ref{EinsteineqB}) gives
\begin{equation} \label{eqB1}
\ddot B_1(t,x) + L \, B_1(t,x) = 0 \; .
\end{equation}
We expand $B_1$ as
\begin{equation} 
\label{seriesB1}
B_1(t,x) = \sum_{n=0}^{\infty} c_n^{(1)}(t) e_n(x) \; .
\end{equation}
The coefficients $c_n^{(1)} \equiv c_n$ satisfy 
\begin{equation}
\label{oscilator}
\ddot{c}_n+\omega_n^2 c_n=0
\end{equation} 
and are given by 
\begin{equation}
\label{cn}
c_n(t) = a_n \cos \left( \theta_n(t) \right) \;,
\end{equation}
with
\begin{equation}
\label{thetan}
\theta_n(t) = \omega_n t + \phi_n \, ,
\end{equation}
where the amplitudes $a_n$ and phases $\phi_n$ are determined by the initial conditions. 

\subsection{Second order perturbations} 

At the second order the equations~(\ref{Einsteineq}) reduce to
\begin{subequations}
\begin{equation} \label{eqB2}
\ddot B_2(t,x) + L \, B_2(t,x) = {40 \over \sin^2 x} B_1^2(t,x) =: S^{(2)}\; ,
\end{equation}
\begin{equation}
A_2' (t,x) = \frac{\nu'(x)}{\nu(x)} A_2 (t,x) - 2 \mu(x) \nu(x) \left( B_1'^2(t,x)+\dot{B}_1^2(t,x) \right) - \frac{16}{\sin^2 x} \mu(x) \nu(x) B_1^2(t,x) \, ,  
\end{equation}
\begin{equation}
\delta_2' (t,x) = - 2 \mu(x) \nu(x) \left( B_1'^2(t,x)+\dot{B}_1^2(t,x) \right) \, .
\end{equation}
\end{subequations}
The equations for the metric functions can be easily integrated to yield:
\begin{equation}
\label{A2integral}
A_2(t,x) = -2 \nu(x) \int_0^x \mu(y) \left( B_1'^2(t,y) + \dot{B}_1^2(t,y) + {8 \over \sin^2 y} B_1^2(t,y) \right) \, dy \; ,
\end{equation}
\begin{equation}
\label{delata2integral}
\delta_2(t,x) = -2 \int_0^x \mu(y)\nu(y) \left( B_1'^2(t,y)+\dot{B}_1^2(t,y) \right) \, dy \; .
\end{equation}
If we expand $B_2$ in terms of eigen functions of (\ref{eigenEq}) 
\begin{equation}
B_2(t,x) = \sum_{n=0}^{\infty} c_n^{(2)}(t) e_n(x) \; .
\end{equation} 
then equation~\eqref{eqB2} reduces to infinite set of equations for the coefficients $c_n^{(2)}$
\begin{equation} \label{eqc2}
\ddot{c}_n^{(2)}+\omega_n^2 c_n^{(2)} = \scalar{S^{(2)}}{e_n} := S^{(2)}_n = 40 \sum_{i} \sum_{j} K_{ijn} \, c_i(t) c_j(t) \, ,
\end{equation}
where 
\begin{equation}
K_{ijn} = \int_0^{\frac{\pi}{2}} \frac{\mu(x)}{\sin^2 x} e_i(x) e_j(x) e_n(x) \, dx
\end{equation}
is the first example of integrals of product of AdS linear eigen modes and some weights that we call (eigen mode) interaction coefficients and that will be frequently encountered in the following sections (for clarity we will list all their definitions while considering the third order equations). In general, at each order of perturbation expansion we will get a forced harmonic oscillator equation
\begin{equation} \label{forcec_oscilator_k}
\ddot{c}_n^{(k)}(t)+\omega_n^2 c_n^{(k)}(t) = S^{(k)}_n  \, ,
\end{equation}
where the source $S^{(k)}_n$ is a sum of products of the first order coefficients $c_i$ multiplied by some eigen mode interaction coefficients. Multiplication of $c_i$ coefficient is governed by the formula 
\begin{equation} 
\cos\theta_i \, \cos\theta_j = \frac{1}{2} \left[ \cos \left( \theta_i + \theta_j \right) + \cos \left( \theta_i - \theta_j \right) \right] \, .
\end{equation}
Whenever, in the result of such multiplication, the source term $S^{(k)}_n$ in (\ref{forcec_oscilator_k}) acquires a resonant term i.e. a term of the  form $\mathcal{A}\cos(\omega_n t + \phi)$, such term results in a term that grows linearly with time $t$ in the solution $c_n^{(k)}$ (called a secular term):
\begin{equation}
\ddot{c}_n^{(k)}(t)+\omega_n^2 c_n^{(k)}(t) = \mathcal{A}\cos(\omega_n t + \phi) + ... \quad \Longrightarrow \quad c_n^{(k)}(t) = \frac{\mathcal{A}}{2 \omega_n} t \sin(\omega_n t + \phi) + ... \, .  
\end{equation}
Thus the presence of resonant terms in the source invalidates na\"ive perturbation expansion at the $\varepsilon^{(k-1)} t$ time scale and such resonant terms dominate the dynamics of the coefficient $c_n^{(k)}$. Craps, Evnin \& Vanhoof (2014)~\cite{cev_JHEP1410} showed how to resum such secular terms, arising from resonant terms in the source, in a systematic way based on renormalization group (RG) method (the reader is strongly encouraged to consult Secs. 1 and 2 of this excellent paper and the references therein to get a broader perspective on long-time effects of small perturbations in Hamiltonian systems and a detailed description of their RG framework). In the case of the massless scalar field studied in \cite{br_PRL107} the resonant terms appear at the third order. As the result of resummation of the resulting secular terms the first order amplitudes and phases are replaced by the slowly varying functions of the "slow" time $\tau=\varepsilon^2 t$:
\begin{align}
c_n & \longrightarrow C_n(\varepsilon^2 t), \quad C_n(0) = c_n
\\
\phi_n & \longrightarrow \Phi_n(\varepsilon^2 t), \quad \Phi_n(0) = \phi_n
\end{align}
Thus it is crucial, at each order of perturbation expansion (\ref{series}), to identify all resonant terms in the source $S^{(k)}_n$. At second order there are no secular terms because the coefficients $K_{ijk}$ vanish for the values of indices $i,j,k$ satisfying the resonance condition, what we prove in Appendix A. 
The solution to~\eqref{eqc2} is given by
\begin{align}
c_n^{(2)} & =  D_1 \sin (\omega_n t) + D_2 \cos (\omega_n t) + \frac{40}{\omega_n} \sum_{i=0}^\infty \sum_{j=0}^\infty K_{ijn}  
\nonumber \\
& \times \left( 
  \sin (\omega_n t) \int_0^t c_i(t') c_j(t') \cos (\omega_n t') \, dt'
- \cos (\omega_n t) \int_0^t c_i(t') c_j(t') \sin (\omega_n t') \, dt'  
\right) \, ,
\end{align}
where $D_1, D_2 = const.$ Zero initial conditions, $B_2(0,x) = 0 = \dot B_2(0,x)$, imply $D_1 = D_2 = 0$. 

\subsection{Third order perturbations and the renormalization flow equations} 

At the third order the equation ~\eqref{EinsteineqB} reduce to
\begin{align}
\ddot{B}_3 + L B_3 &= 2 \left(A_2 - \delta_2 \right) \ddot{B_1} +\left( A_2'-\delta_2' \right) B_1' + \left( \dot{A}_2-\dot{\delta}_2 \right) \dot{B}_1 
\nonumber \\
&- {112\over \sin^2 x} B_1^3 + {80\over \sin^2 x} B_1 B_2 + {8\over \sin^2 x} A_2 B_1 =: S^{(3)}
\label{eqB3}
\end{align}
Expanding $B_3$ into eigenmodes
\begin{equation}
B_3(t,x) = \sum_{n=0}^{\infty} c_n^{(3)}(t) e_n(x) \; .
\end{equation} 
and then projecting \eqref{eqB3} onto the eigen mode basis we get
\begin{equation}
\ddot{c}_l^{(3)}+\omega_l^2 c_l^{(3)} = \scalar{S^{(3)}}{e_l} =: S^{(3)}_l \; ,
\label{c3n}
\end{equation}
with
\begin{align}\label{eq:Sl}
S^{(3)}_l&= 
  2 \scalar{A_2 \ddot B_1}{e_l} 
- 2 \scalar{\delta_2 \ddot B_1}{e_l} 
+ \scalar{\left( A_2' - \delta_2' \right) B_1'}{e_l} 
+ \scalar{\dot A_2 \dot B_1}{e_l} 
- \scalar{\dot \delta_2 \dot B_1}{e_l} 
\nonumber \\
& 
-112 \scalar{{1 \over \sin^2 x} B_1^3}{e_l} 
+ 80 \scalar{{1\over \sin^2 x} B_2 B_1}{e_l} 
+ 8 \scalar{{1\over \sin^2 x} A_2 B_1}{e_l} \, .
\end{align}
This expression strongly resembles an analogical equation of Craps, Evnin \& Vanhoof (2014)~\cite{cev_JHEP1410}, obtained for the massless scalar field. However, it contains three additional terms which are not present in massless, spherically symmetric case, namely $\scalar{{1\over \sin^2 x} B_1^3}{e_l}$, $\scalar{{1\over \sin^2 x} B_1 B_2}{e_l}$, $\scalar{{1\over \sin^2 x} B_1 A_2}{e_l}$. After long and tedious calculation (the details are given in Appendix B) the source term in (\ref{c3n}) can be put in the form:
\begin{align}
& S^{(3)}_l
\nonumber\\
= & \sum_{i,k} a_i^2 a_k \left( - H_{iikl} - 2 \omega_i^2 M_{kli} + 2 \omega_k^2 X_{iikl} - 8 \tilde{X}_{iikl} + 4 \omega_i^2 \omega_k^2 W_{klii} - 16 \omega_i^2 \tilde{W}_{klii} \right) \cos \left( \theta_k \right)
\nonumber\\
- & \frac{1}{2} \sum_{i,j} a_i a_j a_l \omega_l 
\left[ \left( \omega_i \omega_j P_{ijl} + B_{ijl} \right) \left( 2 \omega_l + \omega_j - \omega_i \right) \cos \left( \theta_i - \theta_j - \theta_l \right) \times 2 \right.
\nonumber\\
& \hspace{20mm} - \left( \omega_i \omega_j P_{ijl} - B_{ijl} \right) \left( 2 \omega_l - \omega_j - \omega_i \right) \cos \left( \theta_i + \theta_j - \theta_l \right)
\nonumber\\
& \hspace{20mm} \left. - \left( \omega_i \omega_j P_{ijl} - B_{ijl} \right) \left( 2 \omega_l + \omega_j + \omega_i \right) \cos \left( \theta_i + \theta_j + \theta_l \right) \right]
\nonumber\\
+ & \sum_{i,j,k} a_i a_j a_k \cos \left( \theta_i + \theta_j - \theta_k \right) \times \left\{  - \frac{\omega_j}{\omega_j + \omega_i} \left( 8 \tilde{X}_{ijkl} + H_{ikjl} - 2 \omega_k^2 X_{ijkl} \right) \right.
\nonumber\\ & 
+ [j \neq k] \frac{\omega_j}{\omega_k - \omega_j} \left( 8 \tilde{X}_{kjil} + H_{kjil} - 2 \omega_i^2 X_{kjil} \right) 
+ [i \neq k] \frac{\omega_k}{\omega_i - \omega_k} \left( 8 \tilde{X}_{ijkl} + H_{ikjl} - 2 \omega_j^2 X_{ijkl} \right)
\nonumber\\&
- \omega_j \omega_k X_{ijkl} \times 2 + \omega_i \omega_j X_{kijl} - 4 \tilde{X}_{kijl} - 4 \tilde{X}_{ijkl} \times 2 - 28 G_{ijkl} \times 3 
\nonumber\\&
+ [i \neq l] \frac{\omega_i \left( 2 \omega_i + \omega_j - \omega_k \right)}{ 2 \left(\omega_l^2 - \omega_i^2\right)} Z^+_{kjil} 
+ [j \neq l] \frac{\omega_j \left( 2 \omega_j + \omega_i - \omega_k \right)}{ 2 \left(\omega_l^2 - \omega_j^2\right)} Z^+_{ikjl} 
\nonumber\\&
\left. 
- [k \neq l] \frac{\omega_k \left( 2 \omega_k - \omega_i - \omega_j \right)}{ 2 \left(\omega_l^2 - \omega_k^2\right)} Z^-_{ijkl} \right\}
\nonumber\\
+ & \sum_{i,j,k} a_i a_j a_k \cos \left( \theta_i + \theta_j + \theta_k \right) \times \left\{  - \frac{\omega_j}{\omega_j + \omega_i} \left( 8 \tilde{X}_{ijkl} + H_{ikjl} - 2 \omega_k^2 X_{ijkl} \right) \right.
\nonumber\\&
\left. 
- \omega_j \omega_k X_{ijkl} - 4 \tilde{X}_{ijkl} - 28 G_{ijkl}
- [k \neq l] \frac{\omega_k \left( 2 \omega_k + \omega_i + \omega_j \right)}{ 2 \left(\omega_l^2 - \omega_k^2\right)} Z^-_{ijkl} \right\}
\nonumber\\
+ & 80 \scalar{\frac{1}{\sin^2 x} B_2 B_1}{e_l} \, ,
\label{completeS3l}
\end{align}
where the interaction coefficients are defined as
\begin{subequations}
\label{eq:coeffs}
\begin{align}
\label{Xijkl}
X_{ijkl}&=\int_{0}^{\frac{\pi}{2}}\text{d}x\,e'_{i}(x)e_{j}(x)e_{k}(x)e_{l}(x)(\mu(x))^{2}\nu(x), 
\\
\label{Yijkl}
Y_{ijkl}&=\int_{0}^{\frac{\pi}{2}}\text{d}x\,e'_{i}(x)e_{j}(x)e'_{k}(x)e'_{l}(x)(\mu(x))^{2}\nu(x), 
\\
\label{Hijkl}
H_{ijkl}&=\int_{0}^{\frac{\pi}{2}}\text{d}x\,e'_{i}(x)e_{j}(x)e'_{k}(x)e_{l}(x)(\mu(x))^{2}\nu'(x) 
\\
\label{Zijkl}
Z^{\pm}_{ijkl}&=\omega_{i}\omega_{j}(X_{klij}-X_{lkij})\pm(Y_{klij}-Y_{lkij}),
\\
\label{Wijkl}
W_{ijkl} &= \int_{0}^{\frac{\pi}{2}}\text{d}x\,e_{i}(x)e_{j}(x)\mu(x)\nu(x)\int_{0}^{x}\text{d}y \, \mu(y) \, e_k(y)e_l(y) \, , 
\\
\label{WSijkk}
\bar{W}_{ijkl}&=\int_{0}^{\frac{\pi}{2}}\text{d}x\,e_{i}'(x)e_{j}'(x)\mu(x)\nu(x)\int_{0}^{x}\text{d}y \, \mu(y) \, e_k(y)e_l(y) \, , 
\\
\label{Vij}
V_{ij}&=\int_{0}^{\frac{\pi}{2}}\text{d}x\,e_{i}(x)e_{j}(x)\mu(x)\nu(x) \, , 
\\
\label{Aij}
A_{ij}&=\int_{0}^{\frac{\pi}{2}}\text{d}x\,e'_{i}(x)e'_{j}(x)\mu(x)\nu(x) \, .
\\
P_{ijk} &= V_{ij} - W_{ijkk}
\\
B_{ijk} &= A_{ij} - \bar{W}_{ijkk}
\\
M_{ijk}&=\int_{0}^{\frac{\pi}{2}}\text{d}x\,e'_{i}(x)e_{j}(x)\mu(x)\nu'(x)\int_{0}^{x}\text{d}y(e_{k}(y))^{2}\mu(y), 
\\
\label{Kijk}
K_{ijk}&=\int_{0}^{\frac{\pi}{2}}\text{d}x\, {1\over \sin^2 x}e_{i}(x)e_{j}(x)e_{k}(x)\mu(x),
\\
\label{Gijkl}
G_{ijkl}&=\int_{0}^{\frac{\pi}{2}}\text{d}x\, {1\over \sin^2 x}e_{i}(x)e_{j}(x)e_{k}(x)e_{l}(x)\mu(x),
\\
\tilde{X}_{ijkl}&=\int_{0}^{\frac{\pi}{2}}\text{d}x\, {1\over \sin^2 x}e'_{i}(x)e_{j}(x)e_{k}(x)e_{l}(x)(\mu(x))^{2}\nu(x),
\\
\tilde{W}_{ijkl}&=\int_{0}^{\frac{\pi}{2}}\text{d}x\,{1\over \sin^2 x}e_{i}(x)e_{j}(x)\mu(x)\nu(x)\int_{0}^{x}\text{d}y \, \mu(y) \, e_k(y)e_l(y) \, ,
\end{align}
\end{subequations}
and we used a convenient notation:
\begin{equation}
[condition] = \left\{ \begin{matrix} 
1 \mbox{ if \textit{condition} is true} 
\\
0 \mbox{ if \textit{condition} is false} 
\end{matrix}\right. \, .
\label{[condition]}
\end{equation}
Now, we are ready to identify resonant terms in the source $S^{(3)}_l$. As discussed by Craps, Evnin \& Vanhoof (2014)~\cite{cev_JHEP1410} these terms dictate the dynamics of the system, in particular they control the flow of (conserved) energy between the modes. The resonant terms in $S^{(3)}_l$ are those with $\cos(\pm \omega_l t + \phi)$ time dependence. Such terms come from the following terms in (\ref{completeS3l}) under the following conditions (cf. (\ref{thetan})) (the reason for the single and double underlining of some terms in two following pages will be explained on page~\pageref{WhyUnderlining}).
\paragraph{$\cos(\theta_k)$ terms:}
\begin{center}
\begin{tabular}{l|l}
\begin{minipage}[t]{0.45\textwidth}
$\omega_k = \omega_l$; this gives 
\\
$[k=l] = [k=l](\underline{[i \neq l]} + \doubleunderline{[i=l]})$ 
\\
and contributes to the $(+,+,-)$ resonance, cf. \cite{cev_JHEP1410}, see below.
\end{minipage}
&
\begin{minipage}[t]{0.45\textwidth}
$\omega_k = - \omega_l$ is never satisfied.
\end{minipage}
\end{tabular}
\end{center}
\paragraph{$\cos(\theta_i - \theta_j - \theta_l)$ terms:}
\begin{center}
\begin{tabular}{l|l}
\begin{minipage}[t]{0.45\textwidth}
$\omega_i - \omega_j - \omega_l = \omega_l$ 
\\
these terms do not contribute because for $\omega_i - \omega_j = 2 \omega_l$ their prefactor $2 \omega_l + \omega_j - \omega_i$ is zero.
\end{minipage}
&
\begin{minipage}[t]{0.45\textwidth}
$\omega_i - \omega_j - \omega_l = - \omega_l$; this gives 
\\
$[i=j] = [i=j](\underline{[i=j \neq l]} + \doubleunderline{[i=j=l]})$ 
\\
and contributes to the $(+,+,-)$ resonance, cf. \cite{cev_JHEP1410}, see below.
\end{minipage}
\end{tabular}
\end{center}
\paragraph{$\cos(\theta_i + \theta_j - \theta_l)$ terms:}
\begin{center}
\begin{tabular}{l|l}
\begin{minipage}[t]{0.45\textwidth}
$\omega_i + \omega_j - \omega_l = \omega_l$ 
\\
these terms do not contribute because for $\omega_i + \omega_j = 2 \omega_l$ their prefactor $2 \omega_l - \omega_i - \omega_j$ is zero.
\end{minipage}
&
\begin{minipage}[t]{0.45\textwidth}
$\omega_i + \omega_j - \omega_l = - \omega_l$ is never satisfied. 
\end{minipage}
\end{tabular}
\end{center}
\paragraph{$\cos(\theta_i + \theta_j + \theta_l)$ terms:}
\begin{center}
$\omega_i + \omega_j + \omega_l = \pm \omega_l$ is never satisfied. 
\end{center}
\paragraph{$\cos(\theta_i + \theta_j - \theta_k)$ terms:}
\begin{center}
\begin{tabular}{l|l}
\begin{minipage}[t]{0.45\textwidth}
$\omega_i + \omega_j - \omega_k = \omega_l$; this gives 
\\
$[i+j=k+l]$
\\
and contributes to the $(+,+,-)$ resonance, cf. \cite{cev_JHEP1410}; its name comes from the $\omega_i + \omega_j - \omega_k = \omega_l$ condition with two '$+$' and one '$-$' on the left hand side of the equation.
\end{minipage}
&
\begin{minipage}[t]{0.45\textwidth}
$\omega_i + \omega_j - \omega_k = - \omega_l$; this gives 
\\
$[k=i+j+l+6]$,
\\
this is the $(+,-,-)$ resonance, cf. \cite{cev_JHEP1410}, as $\omega_k - \omega_i - \omega_j = \omega_l$; its name comes from one '$+$' and two '$-$' on the left hand side of the equation.
\end{minipage}
\end{tabular}
\end{center}
\paragraph{$\cos(\theta_i + \theta_j + \theta_k)$ terms:}
\begin{center}
\begin{tabular}{l|l}
\begin{minipage}[t]{0.45\textwidth}
$\omega_i + \omega_j + \omega_k = \omega_l$; this gives 
\\
$[i+j+k+6=l]$ 
\\
this is the $(+,+,+)$ resonance, cf. \cite{cev_JHEP1410}; its name comes from three '$+$' on the left hand side of the equation.
\end{minipage}
&
\begin{minipage}[t]{0.45\textwidth}
$\omega_i + \omega_j + \omega_k = - \omega_l$ is never satisfied. 
\end{minipage}
\end{tabular}
\end{center}
It is also shown in Appendix~B that
\begin{align}
& 80 \scalar{\frac{1}{\sin^2 x} B_2 B_1}{e_l} = -800 \sum_{0 < i,j,k} a_i a_j a_k \cos \left( \theta_i + \theta_j - \theta_k \right) 
\nonumber\\
& \times \left\{ 
\sum_{0 < m} \frac{K_{jkm} K_{ilm}}{\left( \omega_j - \omega_k\right)^2 - \omega_m^2} 
+ \sum_{0 < m} \frac{K_{ikm} K_{jlm}}{\left( \omega_i - \omega_k\right)^2 - \omega_m^2} 
+ \sum_{0 < m} \frac{K_{ijm} K_{klm}}{\left( \omega_i + \omega_j\right)^2 - \omega_m^2}
\right\}
\nonumber\\
& + \mbox{non-resonant terms } \, 
\label{B2B1oversin}
\end{align}
and there is no contribution from (\ref{B2B1oversin}) neither to $(+,+,+)$ nor $(+,-,-)$ resonances. The sums in (\ref{B2B1oversin}) are understood in such way that there is no contribution whenever numerators are zero, thus there is no problem with divisions by zero as $K_{ijk} \equiv 0$ for any permutation of indices in the inequality $k > i+j+2$. Finally $S^{(3)}_l$ takes the following form:
\begin{align}
S^{(3)}_l & = \doubleunderline{a_l^3 T_l \cos\left( \theta_l \right)} 
+ \underline{\sum_{0 < i \neq l} a_l a_i^2 R_{il} \cos\left( \theta_l \right)}
+ \sum_{\scriptsize{\begin{matrix} 0 < i,j,k \\ i+j = k+l \\ i \neq l \neq j \end{matrix}}} a_i a_j a_k S_{ijkl} \cos\left( \theta_i + \theta_j - \theta_k \right)
\nonumber\\
& + \sum_{\scriptsize{\begin{matrix} 0 < i,j,k \\ k = i+j+l+6 \end{matrix}}} a_i a_j a_k U_{kijl} \cos\left( \theta_k - \theta_i - \theta_j \right)
+ \sum_{\scriptsize{\begin{matrix} 0 < i,j,k \\ i+j+k+6 = l \end{matrix}}} a_i a_j a_k Q_{ijkl} \cos\left( \theta_i + \theta_j + \theta_k \right)
\nonumber\\
& + \mbox{non-resonant terms .}
\label{S3lfinal}
\end{align}
\label{WhyUnderlining}
To identify contributions to $T_l$, $R_{il}$ and $S_{ijkl}$ in (\ref{S3lfinal}) we note following identities to be used under sums in (\ref{completeS3l}) (contributions to $R_{il}$ and $T_l$ are marked with single and double underlining here an in the text between eq.~(\ref{[condition]}) and eq.~(\ref{B2B1oversin}))
\begin{align}
[i+j=k+l][j \neq k] &= [i+j=k+l][i \neq l \neq j] + \underline{[j=l][i = k \neq l]} = [i+j=k+l][i \neq l]  \, ,
\\
[i+j=k+l][i \neq k] &= [i+j=k+l][i \neq l \neq j] + \underline{[i=l][j = k \neq l]} = [i+j=k+l][j \neq l] \, ,
\\
[i+j=k+l][k \neq l] &= [i+j=k+l][i \neq l \neq j] + \underline{[j=l][i = k \neq l]} + \underline{[i=l][j = k \neq l]} \, 
\end{align}
and
\begin{align}
& [i+j=k+l] 
\nonumber\\
= & [i+j=k+l][i \neq l \neq j] + \underline{[j=l][i=k \neq l]} + \underline{[i=l][j=k \neq l]} + \doubleunderline{[i=j=k=l]} \, .
\end{align}
This leads to 
\begin{align}
T_l & = -\frac{3}{2} H_{llll} + 2 \omega_l^2 X_{llll} - 24 \tilde{X}_{llll} - 2 \omega_l^2 M_{lll} - 16 \omega_l^2 \tilde{W}_{llll} + 4 \omega_l^4 W_{llll} - 2 \omega_l^2 \left( \omega_l^2 P_{lll} + B_{lll} \right)
\nonumber \\
& - 84 G_{llll} - 800 \sum_{0 \leq m \leq 2l+2} \left( \frac{1}{4 \omega_l^2 - \omega_m^2} - \frac{2}{\omega_m^2} \right) \left(K_{llm}\right)^2 \, ,
\label{Tl}
\end{align}
\begin{align}
R_{il} & = 2\left(\frac{\omega_{i}^{2}}{\omega_{l}^{2}-\omega_{i}^{2}}\right)\left(H_{liil}-2\omega_{i}^{2}X_{liil}+8 \tilde{X}_{liil} \right) 
\nonumber \\
& -2\left(\frac{\omega_{l}^{2}}{\omega_{l}^{2}-\omega_{i}^{2}}\right)\left(H_{ilil}-2\omega_{i}^{2}X_{ilil}+8 \tilde{X}_{ilil}\right) 
\nonumber \\
& -2\omega_{i}^{2}X_{liil} -24 \tilde{X}_{ilil} -8 \tilde{X}_{liil} 
\nonumber \\
& -\left(H_{iill}+2\omega_{i}^{2}M_{lli}\right)+2\omega_{l}^{2}\left(X_{iill}+2\omega_{i}^{2}W_{lli}\right)-16\omega_{i}^{2} \tilde{W}_{lli}-2\omega_{l}^{2}\left(\omega_{i}^{2}P_{iil}+B_{iil}\right) 
\nonumber \\
& +2\left(\frac{\omega_{i}^{2}}{\omega_{l}^{2}-\omega_{i}^{2}}\right)\left(Y_{illi}-Y_{lili}+\omega_{l}^{2}(X_{illi}-X_{lili})\right) - 168G_{ilil} 
\nonumber \\
& -1600 \sum_{\scriptsize{\begin{matrix}m=0\\ i-l\neq  \pm (m+3)\end{matrix}}}^{i+l+2} \left( \frac{1}{(\omega_i-\omega_l)^2-\omega_m^2}+\frac{1}{(\omega_i+\omega_l)^2-\omega_m^2} \right) (K_{ilm})^2 
\nonumber \\
& +1600 \sum_{m=0}^{\scriptsize{\begin{matrix}m<2i+3\\ m<2l+3 \end{matrix}}} \frac{1}{\omega_m^2} K_{iim} K_{llm} \, , 
\label{Ril}
\end{align}
\begin{align}
S_{ijkl}=&-\frac{1}{2}H_{ijkl}\omega_{j}\left(\frac{1}{\omega_{j}+\omega_{i}}+\frac{1}{\omega_{j}-\omega_{k}}\right)-\frac{1}{2}H_{jkil}\omega_{k}\left(\frac{1}{\omega_{k}-\omega_{i}}+\frac{1}{\omega_{k}-\omega_{j}}\right) \nonumber \\
&-\frac{1}{2}H_{kijl}\omega_{i}\left(\frac{1}{\omega_{i}+\omega_{j}}+\frac{1}{\omega_{i}-\omega_{k}}\right)+X_{kijl}\omega_{i}\omega_{j}\left(\frac{\omega_{j}}{\omega_{i}-\omega_{k}}+\frac{\omega_{i}}{\omega_{j}-\omega_{k}}+1\right) \nonumber \\
&+X_{ijkl}\omega_{j}\omega_{k}\left(\frac{\omega_{k}}{\omega_{j}+\omega_{i}}+\frac{\omega_{j}}{\omega_{k}-\omega_{i}}-1\right)+X_{jkil}\omega_{k}\omega_{i}\left(\frac{\omega_{k}}{\omega_{i}+\omega_{j}}+\frac{\omega_{i}}{\omega_{k}-\omega_{j}}-1\right) \nonumber \\
&+\frac{1}{2}\left(\frac{\omega_{k}}{\omega_{i}+\omega_{j}}\right)Z^{-}_{ijkl}+\frac{1}{2}\left(\frac{\omega_{i}}{\omega_{j}-\omega_{k}}\right)Z^{+}_{jkil}+\frac{1}{2}\left(\frac{\omega_{j}}{\omega_{i}-\omega_{k}}\right)Z^{+}_{kijl} \nonumber \\
&-4 \tilde{X}_{ijkl}\left(1+\frac{\omega_{j}}{\omega_{i}+\omega_{j}}+\frac{\omega_{k}}{\omega_{k}-\omega_{i}}\right)-4 \tilde{X}_{jkil}\left(1+\frac{\omega_{i}}{\omega_{i}+\omega_{j}}+\frac{\omega_{k}}{\omega_{k}-\omega_{j}}\right) \nonumber \\
&-4 \tilde{X}_{kijl}\left(1+\frac{\omega_{i}}{\omega_{i}-\omega_{k}}+\frac{\omega_{j}}{\omega_{j}-\omega_{k}}\right)\nonumber \\
&-84G_{ijkl}-800 \sum_{m=0}^{i+j+2} \frac{1}{(\omega_i+\omega_j)^2-\omega_m^2} K_{ijm}K_{mkl} \nonumber \\
&-800 \sum_{\scriptsize{\begin{matrix}m=0\\ i-k\neq \pm (m+3)\end{matrix}}}^{\scriptsize{\begin{matrix}m<i+k+3\\ m<l+j+3\end{matrix}}} \frac{1}{(\omega_i-\omega_k)^2-\omega_m^2} K_{ikm}K_{mjl}  \nonumber \\
&-800 \sum_{\scriptsize{\begin{matrix}m=0\\ j-k\neq \pm (m+3)\end{matrix}}}^{\scriptsize{\begin{matrix}m<j+k+3\\ m<l+i+3\end{matrix}}} \frac{1}{(\omega_j-\omega_k)^2-\omega_m^2} K_{jkm}K_{mil} \, , 
\label{Sijkl}
\end{align}
where $S_{ijkl}$ is taken to be symmetric in its first two indices and it is understood that on both sides of (\ref{Sijkl}) the condition $[i+j=k+l][i \neq l \neq j]$ holds. 

One can show that $U_{ijkl}$ and $Q_{ijkl}$ contain no contribution from scalar products $\scalar{{1\over \sin^2 x} B_1^3}{e_l}$, $\scalar{{1\over \sin^2 x}B_1 B_2}{e_l}$ (for details see Appendix B). For the $U_{ijkl}$ terms we get:
\begin{align}
& U_{ijkl} \times [i=j+k+l+6] 
\nonumber\\
&= \left[ 
\frac{1}{2}H_{ijkl}\frac{\omega_{j}(2\omega_{j}-\omega_{i}+\omega_{k})}{(\omega_{i}-\omega_{j})(\omega_{j}+\omega_{k})}+\frac{1}{2}H_{jkil}\frac{\omega_{k}(2\omega_{k}-\omega_{i}+\omega_{j})}{(\omega_{i}-\omega_{k})(\omega_{k}+\omega_{j})}+\frac{1}{2}H_{kijl}\frac{\omega_{i}(\omega_{j}+\omega_{k}-2\omega_{i})}{(\omega_{i}-\omega_{j})(\omega_{i}-\omega_{k})} \right.
\nonumber \\
&-X_{ijkl}\,\omega_{j}\omega_{k}\left(\frac{\omega_{k}}{(\omega_{i}-\omega_{j})}+\frac{\omega_{j}}{(\omega_{i}-\omega_{k})}-1\right)+X_{jkil}\,\omega_{i}\omega_{k}\left(\frac{\omega_{k}}{(\omega_{i}-\omega_{j})}+\frac{\omega_{i}}{(\omega_{k}+\omega_{j})}-1\right) \nonumber \\
&+X_{kijl}\,\omega_{i}\omega_{j}\left(\frac{\omega_{i}}{(\omega_{j}+\omega_{k})}+\frac{\omega_{j}}{(\omega_{i}-\omega_{k})}-1\right) \nonumber \\
&-\frac{1}{2}Z^{+}_{ijkl}\frac{\omega_{k}}{(\omega_{i}-\omega_{j})}+\frac{1}{2}Z^{-}_{jkil}\frac{\omega_{i}}{(\omega_{j}+\omega_{k})}-\frac{1}{2}Z^{+}_{kijl}\frac{\omega_{j}}{(\omega_{i}-\omega_{k})} \nonumber \\
&-4 \tilde{X}_{ijkl}\left(1+\frac{\omega_{j}}{(\omega_{j}-\omega_{i})}+\frac{\omega_{k}}{(\omega_{k}-\omega_{i})}\right)-4 \tilde{X}_{jkil}\left(1+\frac{\omega_{i}}{(\omega_{i}-\omega_{j})}+\frac{\omega_{k}}{(\omega_{j}+\omega_{k})}\right) 
\nonumber \\
& 
\left. -4 \tilde{X}_{kijl}\left(1+\frac{\omega_{i}}{(\omega_{i}-\omega_{k})}+\frac{\omega_{j}}{(\omega_{j}+\omega_{k})}\right) \right] \times [i=j+k+l+6] \;.
\label{Uijkl}
\end{align}
For the $Q_{ijkl}$ terms we get:
\begin{align}
& Q_{ijkl} \times [i+j+k+6=l]
\nonumber\\
&= \left[ -\frac{1}{6}H_{ijkl}\frac{\omega_{j}(2\omega_{j}+\omega_{i}+\omega_{k})}{(\omega_{j}+\omega_{i})(\omega_{j}+\omega_{k})}-\frac{1}{6}H_{jkil}\frac{\omega_{k}(2\omega_{k}+\omega_{i}+\omega_{j})}{(\omega_{k}+\omega_{i})(\omega_{k}+\omega_{j})} \right.
\nonumber \\
&-\frac{1}{6}H_{kijl}\frac{\omega_{i}(2\omega_{i}+\omega_{j}+\omega_{k})}{(\omega_{i}+\omega_{j})(\omega_{i}+\omega_{k})}+\frac{1}{3}X_{ijkl}\,\omega_{j}\omega_{k}\left(1+\frac{\omega_{k}}{(\omega_{j}+\omega_{i})}+\frac{\omega_{j}}{(\omega_{k}+\omega_{i})}\right) \nonumber \\
&+\frac{1}{3}X_{jkil}\omega_{i}\omega_{k}\left(1+\frac{\omega_{k}}{(\omega_{i}+\omega_{j})}+\frac{\omega_{i}}{(\omega_{k}+\omega_{j})}\right)+\frac{1}{3}X_{kijl}\,\omega_{i}\omega_{j}\left(1+\frac{\omega_{i}}{(\omega_{j}+\omega_{k})}+\frac{\omega_{j}}{(\omega_{i}+\omega_{k})}\right) \nonumber \\
&-\frac{1}{6}Z^{-}_{ijkl}\frac{\omega_{k}}{(\omega_{i}+\omega_{j})}-\frac{1}{6}Z^{-}_{jkil}\frac{\omega_{i}}{(\omega_{j}+\omega_{k})}-\frac{1}{6}Z^{-}_{kijl}\frac{\omega_{j}}{(\omega_{i}+\omega_{k})} \nonumber \\
&-\frac{4}{3} \tilde{X}_{ijkl} \left(1+\frac{\omega_{j}}{(\omega_{j}+\omega_{i})}+\frac{\omega_{k}}{(\omega_{k}+\omega_{i})}\right)-\frac{4}{3} \tilde{X}_{jkil}\left(1+\frac{\omega_{k}}{(\omega_{k}+\omega_{j})}+\frac{\omega_{i}}{(\omega_{i} +\omega_{j})}\right) \nonumber \\
&\left. -\frac{4}{3} \tilde{X}_{kijl}\left(1+\frac{\omega_{i}}{(\omega_{i}+\omega_{k})}+\frac{\omega_{j}}{(\omega_{j}+\omega_{k})}\right) \right] \times [i+j+k+6=l] \;.
\label{Qijkl}
\end{align}
With the help of identities \footnote{To prove (\ref{Hidentity}) we integrate $(-Y_{klij})$ by parts: 
\[
-Y_{klij} =- \int_{0}^{\pi/2} dx \, e'_k e_l e'_i e'_j \mu^2 \nu = \int_{0}^{\pi/2} dx \, e_j  \left[ \left(\mu e'_k\right) \left(\mu e'_i\right) e_l \nu \right]' 
\]
and use the eigen equation (\ref{eigenEq}) in a form $\left(\mu e'_k\right)' = \frac{8}{\sin^2(x)} \mu e_k - \omega_k^2  e_k$. This cancels all other terms on the RHS of (\ref{Hidentity}) and leaves $H_{ijkl}$. Similarly, to prove (\ref{Midentity}) we integrate $(-X_{ijkk})$ by parts: 
\[
-X_{ijkk} =- \int_{0}^{\pi/2} dx \, e'_i e_j \mu \nu \left(\mu  e^2_k \right) = \int_{0}^{\pi/2} dx \left[ \left(\mu e'_i \right) e_j \nu\right]' \int_0^x dy\, \mu e^2_k 
\]
and use the eigen equation (\ref{eigenEq}). This cancels all other terms on the RHS of (\ref{Midentity}) and leaves $M_{ijk}$.}
\begin{subequations}
\label{HMidentities}
\begin{align}
H_{ijkl} & = \omega_i^2 X_{klij} - 8\tilde{X}_{klij} + \omega_k^2 X_{ijkl} - 8\tilde{X}_{ijkl} - Y_{klij} - Y_{ijkl} \;, 
\label{Hidentity}
\\
M_{ijk} & =\omega_i^2 W_{ijk} - 8\tilde{W}_{ijkk} - X_{ijkk} + B_{ijk} - A_{ij} \;,
\label{Midentity}
\end{align}
\end{subequations}
expressions~\eqref{Uijkl},~\eqref{Qijkl} can be simplified to yield:
\begin{align} \label{Uijklfinal}
& U_{ijkl}\times [i=j+k+l+6] 
\nonumber\\
 &= \left[ \frac{1}{2} \left(\frac{1}{\omega_{i}-\omega_{j}}-\frac{1}{\omega_{k}-\omega_{i}}-\frac{1}{\omega_{j}+\omega_{k}}\right) (\omega_{i} \omega_{j} \omega_{k} X_{lijk}+\omega_{l} Y_{iljk}) \right.
\nonumber \\
&+\frac{1}{2} \left(\frac{1}{\omega_{i}-\omega_{j}}+\frac{1}{\omega_{k}-\omega_{i}}+\frac{1}{\omega_{j}+\omega_{k}}\right) (\omega_{i} \omega_{j} \omega_{l} X_{kijl}+\omega_{k} Y_{ikjl}) \nonumber \\
&+\frac{1}{2} \left(-\frac{1}{\omega_{i}-\omega_{j}}-\frac{1}{\omega_{k}-\omega_{i}}+\frac{1}{\omega_{j}+\omega_{k}}\right) (\omega_{i} \omega_{k} \omega_{l} X_{jikl}+\omega_{j} Y_{ijkl}) \nonumber \\
&\left. +\frac{1}{2} \left(\frac{1}{\omega_{i}-\omega_{j}}-\frac{1}{\omega_{k}-\omega_{i}}+\frac{1}{\omega_{j}+\omega_{k}}\right) (\omega_{j} \omega_{k} \omega_{l} X_{ijkl}+\omega_{i} Y_{jikl}) \right] \times [i=j+k+l+6] \; ,
\end{align}
\begin{align} \label{Qijklfinal}
& Q_{ijkl} \times [i+j+k+6=l]
\nonumber\\
 & =\left[\frac{1}{6} \left(\frac{1}{\omega_{i}+\omega_{j}}+\frac{1}{\omega_{i}+\omega_{k}}+\frac{1}{\omega_{j}+\omega_{k}}\right) (\omega_{i} \omega_{j} \omega_{k} X_{lijk}+\omega_{l} Y_{iljk}) \right.\nonumber \\
&+\frac{1}{6} \left(-\frac{1}{\omega_{i}+\omega_{j}}+\frac{1}{\omega_{i}+\omega_{k}}+\frac{1}{\omega_{j}+\omega_{k}}\right) (\omega_{i} \omega_{j} \omega_{l} X_{kijl}+\omega_{k} Y_{ikjl}) \nonumber \\
&+\frac{1}{6} \left(\frac{1}{\omega_{i}+\omega_{j}}-\frac{1}{\omega_{i}+\omega_{k}}+\frac{1}{\omega_{j}+\omega_{k}}\right) (\omega_{i} \omega_{k} \omega_{l} X_{jikl}+\omega_{j} Y_{ijkl}) \nonumber \\
&\left.+\frac{1}{6} \left(\frac{1}{\omega_{i}+\omega_{j}}+\frac{1}{\omega_{i}+\omega_{k}}-\frac{1}{\omega_{j}+\omega_{k}}\right) (\omega_{j} \omega_{k} \omega_{l} X_{ijkl}+\omega_{i} Y_{jikl}) \right] \times [i+j+k+6=l] \;.
\end{align}
Our numerical results show that both these expressions vanish, like in case of Einstein equations with a massless scalar field~\cite{cev_JHEP1410}. Similarly, using (\ref{Hidentity},\ref{Midentity}) expressions~\eqref{Tl},~\eqref{Ril},~\eqref{Sijkl} can be simplified to yield:
\begin{align} 
\label{Tlfinal}
T_l&=\omega_{l}^2 X_{llll}+3 Y_{llll}+4\omega_{l}^4 W_{llll}+4\omega_{l}^2 \bar{W}_{llll}-2\omega_{l}^2 \left(A_{ll}+\omega_{l}^2 V_{ll}\right) \nonumber \\
& -84 G_{llll} - 800 \sum_{m=0}^{2l+2} \left( \frac{1}{4\omega_l^2-\omega_m^2}-\frac{2}{\omega_m^2} \right)(K_{llm})^2 \; ,
\end{align}
\begin{align} \label{Rilfinal}
R_{il}&=
\frac{\left(\omega_{i}^2+\omega_{l}^2\right) \left(\omega_{l}^2 X_{illi}-\omega_{i}^2 X_{liil}\right)}{ \left(\omega_{l}^2-\omega_{i}^2\right)}+\frac{4 \left(\omega_{l}^2 Y_{ilil}-\omega_{i}^2 Y_{lili}\right)}{\omega_{l}^2-\omega_{i}^2} \nonumber \\
&+2\frac{\omega_{i}^2 \omega_{l}^2 (X_{illi}-X_{lili})}{\omega_{l}^2-\omega_{i}^2}+Y_{iill}+Y_{llii}+2\omega_{i}^2 \omega_{l}^2 (W_{iill}+W_{llii}) \nonumber \\
&+2\omega_{i}^2 \bar{W}_{llii}+2\omega_{l}^2 \bar{W}_{iill}-2\omega_{l}^2 \left(A_{ii}+\omega_{i}^2 V_{ii}\right) - 168G_{ilil} \nonumber \\
&-1600 \sum_{\scriptsize{\begin{matrix}m=0\\ i-l\neq  \pm (m+3)\end{matrix}}}^{i+l+2} \left( \frac{1}{(\omega_i-\omega_l)^2-\omega_m^2}+\frac{1}{(\omega_i+\omega_l)^2-\omega_m^2} \right) (K_{ilm})^2  \nonumber \\
&+1600 \sum_{m=0}^{\scriptsize{\begin{matrix}m<2i+3\\ m<2l+3 \end{matrix}}} \frac{1}{\omega_m^2} K_{iim} K_{mll} \;,  
\end{align}
\begin{align} \label{Sijklfinal}
S_{ijkl}=&-\frac{1}{2} \left(\frac{1}{\omega_{i}+\omega_{j}}+\frac{1}{\omega_{i}-\omega_{k}}+\frac{1}{\omega_{j}-\omega_{k}}\right) (\omega_{i} \omega_{j} \omega_{k} X_{lijk}-\omega_{l} Y_{iljk}) \nonumber \\
&-\frac{1}{2} \left(\frac{1}{\omega_{i}+\omega_{j}}-\frac{1}{\omega_{i}-\omega_{k}}-\frac{1}{\omega_{j}-\omega_{k}}\right)(\omega_{i} \omega_{j} \omega_{l} X_{kijl}-\omega_{k} Y_{ikjl}) \nonumber \\
&-\frac{1}{2} \left(\frac{1}{\omega_{i}+\omega_{j}}-\frac{1}{\omega_{i}-\omega_{k}}+\frac{1}{\omega_{j}-\omega_{k}}\right) (\omega_{i} \omega_{k} \omega_{l} X_{jikl}-\omega_{j} Y_{ijkl}) \nonumber \\
&-\frac{1}{2} \left(\frac{1}{\omega_{i}+\omega_{j}}+\frac{1}{\omega_{i}-\omega_{k}}-\frac{1}{\omega_{j}-\omega_{k}}\right)(\omega_{j} \omega_{k} \omega_{l} X_{ijkl}-\omega_{i} Y_{jikl}) \nonumber \\
&-84G_{ijkl}-800 \sum_{m=0}^{i+j+2} \frac{1}{(\omega_i+\omega_j)^2-\omega_m^2} K_{ijm}K_{mkl} \nonumber \\
&-800 \sum_{\scriptsize{\begin{matrix}m=0\\ i-k\neq \pm (m+3)\end{matrix}}}^{\scriptsize{\begin{matrix}m<i+k+3\\ m<l+j+3\end{matrix}}} \frac{1}{(\omega_i-\omega_k)^2-\omega_m^2} K_{ikm}K_{mjl}  \nonumber \\
&-800 \sum_{\scriptsize{\begin{matrix}m=0\\ j-k\neq \pm (m+3)\end{matrix}}}^{\scriptsize{\begin{matrix}m<j+k+3\\ m<l+i+3\end{matrix}}} \frac{1}{(\omega_j-\omega_k)^2-\omega_m^2} K_{jkm}K_{mil} \;, 
\end{align}
where it is understood that on both sides of (\ref{Sijklfinal}) the condition $[i+j=k+l][i \neq l \neq j]$ holds.
Following Craps, Evnin \& Vanhoof (2014)~\cite{cev_JHEP1410}, we finally obtain renormalization flow equations for non-linear perturbation theory at first non-trivial order
\begin{subequations}
\label{ResonantSystem}
\begin{align}
\label{Cdot}
2 \omega_l \frac{dC_l}{d \tau} &= - \underbrace{\sum_{i,(i\neq l)} \sum_{j,(j\neq l)}}_{l\leq i+j} S_{ij(i+j-l)l} C_i C_j C_{i+j-l}\sin (\Phi_l+\Phi_{i+j-l}-\Phi_i-\Phi_j) \; , 
\\
2\omega_l C_l \frac{d\Phi_l}{d \tau} &= -T_l C_l^3- \sum_{i,(i\neq l)} R_{il} C_i^2 C_l 
\nonumber \\
&\underbrace{\sum_{i,(i\neq l)} \sum_{j,(j\neq l)}}_{l\leq i+j} S_{ij(i+j-l)l} C_i C_j C_{i+j-l}\sin (\Phi_l+\Phi_{i+j-l}-\Phi_i-\Phi_j)\;,
\label{Phidot}
\end{align}
\end{subequations}
where $C_l$ and $\Phi_l$ are the running renormalized amplitudes and phases i.e. the solutions to (\ref{Cdot}, \ref{Phidot}) with initial conditions $C_l(0) = a_l$ and $\Phi_l(0) = \phi_l$ (cf. (\ref{seriesB1}-\ref{thetan})) and the solution resummed up to the first non-trivial order reads:
\begin{equation} 
\label{Bresummed}
B(t,x) = \varepsilon \sum_{n=0}^{\infty} C_n\left(\varepsilon^2 t\right) \, \cos\left(\omega_n t + \Phi_n\left(\varepsilon^2 t\right)\right) \, e_n(x) \; .
\end{equation}

\section{Recurrence relations for the interaction coefficients}
\label{recurrences}

Obtaining interaction coefficients of the resonant system (\ref{Cdot}, \ref{Phidot}) from direct integration of their defining integrals (\ref{eq:coeffs}) is numerically expensive and moreover does not provide much insight into ultraviolet asymptotics of the interaction coefficients that is crucial to understand the asymptotic behavior of solutions of the resonant system. For the massless scalar field model \cite{bbgll_PRL113, cev_JHEP1410} in $d=3$ spatial dimensions the integrals (\ref{eq:coeffs}) can be calculated analytically, providing closed-form formulas for interaction coefficients \cite{gmll_PRD92}. This is possible due to the existence of simplified representation of eigenfunctions and this approach can be generalized to arbitrary odd number of spatial dimensions \cite{m_private1}. However, in the present studies with $d=4$, we are unaware of any such methods of direct analytic evaluation of the interaction coefficients. Thus, to study asymptotic behaviour of solutions of the resonant system, both numerically and analytically, it is useful to provide at least recurrence relations for the interaction coefficients. For the Einstein--massless scalar field system such relations were provided by Craps, Evnin and Vanhoof in \cite{cev_JHEP1510} and in this section we follow their approach. From the definition of eigenfunctions $e_j$ in terms of Jacobi polynomials, cf. (\ref{modes}), and recurrence relations for Jacobi polynomials themselves
\begin{align}
&2(n+1)(n+\alpha+\beta+1)(2n+\alpha+\beta)P_{n+1}^{(\alpha ,\beta)}(x) \nonumber \\
&= -2(n+\alpha)(n+\beta)(2n+\alpha+\beta+2)P_{n-1}^{(\alpha ,\beta)}(x) \nonumber \\ &+(2n+\alpha+\beta+1) \left[ (2n+\alpha+\beta+2)(2n+\alpha+\beta)x+\alpha^2-\beta^2 \right] P_{n}^{(\alpha ,\beta)}(x) \; ,
\end{align}
\begin{align}
&(2n+\alpha+\beta+2)(1-x^2){d\over dx} P_n^{(\alpha ,\beta)}(x) = -2(n+1)(n+\alpha+\beta+1)P_{n+1}^{(\alpha ,\beta)}(x) \nonumber \\
&+(n+\alpha+\beta+1)(\alpha-\beta+(2n+\alpha+\beta+2)x) P_{n}^{(\alpha ,\beta)}(x)
\end{align}
we get
\begin{align} \label{identity1}
\mu\nu'e_n = A_{-}(n)e_n+B(n)e_{n+1}+C(n)e_{n-1} \; ,
\end{align}
\begin{align} \label{identity2}
\mu\nu e_n' = \frac{1}{2} A_{+}(n)e_n + \frac{\omega_n}{2} B(n)e_{n+1} - \frac{\omega_n}{2} C(n)e_{n-1} \; 
\end{align}
with
\begin{equation}
 A_{\pm}(n) = -3 \pm \frac{5}{\omega_n^2 - 1} \, , \qquad B(n) = \frac{\sqrt{(n+1)(n+6)}}{\omega_n + 1} \, , \qquad C(n) = \frac{\sqrt{n(n+5)}}{\omega_n - 1} \, . 
\end{equation}
Now, differentiating (\ref{identity1},\ref{identity2}) and using eigen equation (\ref{eigenEq}) to eliminate $e_n''$ and the identity $\left(\mu \nu'\right)' = -4 \mu \nu$ we get
\begin{align} \label{identity3}
\mu\nu'e_n' = -\frac{16 - 4\omega_n^2 + 3 \omega_n^4}{(\omega_n^2 - 4)(\omega_n^2 - 1)} e_n' + B(n) \frac{\omega_n}{\omega_{n+1}} e_{n+1}' + C(n) \frac{\omega_n}{\omega_{n-1}} e_{n-1}' + \frac{32}{\omega_n^2 - 4} \frac{\mu\nu}{\sin^2 x} e_n\; ,
\end{align}
\begin{align} \label{identity4}
2 \mu\nu e_n = -\frac{3(4 + \omega_n^2)}{(\omega_n^2 - 4)(\omega_n^2 - 1)} e_n' - B(n) \frac{e_{n+1}'}{\omega_{n+1}} + C(n) \frac{e_{n-1}'}{\omega_{n-1}} + \frac{16}{\omega_n^2 - 4} \frac{\mu\nu}{\sin^2 x} e_n \; . 
\end{align}
The identities (\ref{identity1}-\ref{identity4}) are analogous to identities (15-18) in \cite{cev_JHEP1510} for the massless scalar field coupled to Einstein equations. 

\subsection{Recurrence relation for the $X_{mnpq}$ integrals}
Using the identity (\ref{identity2}) in the definition of the $X_{mnpq}$ integral (\ref{Xijkl}), $X_{mnpq}$ can be given in terms of integrals $\chi_{mnpq}$, totally symmetric in their indices:
\begin{equation}
\label{CHImnpq}
\chi_{mnpq} = \int_{0}^{\frac{\pi}{2}} \text{d}x\, \mu(x) e_{m}(x) e_{n}(x) e_{p}(x) e_{q}(x) \, ,
\end{equation}
namely
\begin{equation}
\label{Xmnpq_in CHI}
X_{mnpq} = \frac{1}{2} A_{+}(m) \chi_{mnpq} + \frac{\omega_m}{2} B(m) \chi_{(m+1)npq} - \frac{\omega_m}{2} C(m) \chi_{(m-1)npq} \, .
\end{equation}
Now, to get the recurrence relation for the $\chi_{mnpq}$ integral we consider another auxiliary integral 
\begin{equation}
\label{tildeCHImnpq}
\tilde \chi_{mnpq} = \int_{0}^{\frac{\pi}{2}} \text{d}x\, \mu^2(x) \nu'(x) e_{m}(x) e_{n}(x) e_{p}(x) e_{q}(x) \, .
\end{equation}
In this integral we either (1) use the identity (\ref{identity1}) for $\mu \nu' e_m$, or (2) integrate by parts using $\mu' \nu = d-1$ and the definition (\ref{Xijkl}). Equating the results of these two operations we get (for $d=4$): 
\begin{align}
  & A_{-}(m) \chi_{mnpq} + B(m) \chi_{(m+1)npq} + C(m) \chi_{(m-1)npq} 
\nonumber\\
= & -6  \chi_{mnpq} - X_{mnpq} - X_{npqm} - X_{pqmn} - X_{qmnp} \, .
\label{Xmnpq_aux}
\end{align}
Similarly, using the identity (\ref{identity1}) in sequence for $\mu \nu' e_m$, $\mu \nu' e_n$, $\mu \nu' e_p$, $\mu \nu' e_q$ we get from (\ref{tildeCHImnpq}) a sequence of identities 
\begin{align}
& A_{-}(m) \chi_{mnpq} + B(m) \chi_{(m+1)npq} + C(m) \chi_{(m-1)npq}  
\nonumber\\
= & A_{-}(n) \chi_{npqm} + B(n) \chi_{(n+1)pqm} + C(n) \chi_{(n-1)pqm}  
\nonumber\\
= & A_{-}(p) \chi_{pqmn} + B(p) \chi_{(p+1)qmn} + C(p) \chi_{(p-1)qmn}  
\nonumber\\
= & A_{-}(q) \chi_{qmnp} + B(q) \chi_{(q+1)mnp} + C(q) \chi_{(q-1)mnp}  
\end{align}  
that can be solved for $\chi_{(n+1)pqm}$, $\chi_{(p+1)qmn}$ and $\chi_{(q+1)mnp}$. Then substituting (\ref{Xmnpq_in CHI}) into (\ref{Xmnpq_aux}) we get the recurrence relation for the integral $\chi_{mnpq}$ (totally symmetric in its indices): 
\begin{align}
\chi_{mnpq} = & \frac{1}{(12+m+n+p+q) \sqrt{m (m+5)}} 
\nonumber\\
\times & \left\{ \frac{2 \chi_{(m-1)npq}}{(2m + 3) (2n + 5) (2p + 5) (2q + 5)}
\left[5 \left( 875 + 450(n+p+q) + 25 \left( n^2 + p^2 + q^2 \right) \right.\right.\right.
\nonumber\\
& + 220 (np + nq + pq) + 104 npq + 10 \left(n^2(p+q) + p^2(n+q) + q^2(n+p) \right) 
\nonumber\\
& \left. \left. + 4 \left(n^2pq + np^2q + npq^2 \right) \right) + m (m+4) ( 375 + 200(n+p+q) + 100(np + nq + pq) + 48npq) \right]
\nonumber\\
& + \left[ \frac{m-n-p-q-8}{2 m+3} \sqrt{(m-1) (m+4)} \chi_{(m-2)npq} + \frac{2 (n+3)}{2 n+5} \sqrt{n (n+5)} \chi_{(m-1)(n-1)pq} \right.
\nonumber\\
& \left. \left. + \frac{2 (p+3)}{2 p+5} \sqrt{p (p+5)} \chi_{(m-1)n(p-1)q} + \frac{2 (q+3)}{2 q+5} \sqrt{q (q+5)} \chi_{(m-1)np(q-1)} \right] (2m + 5) \right\} 
\end{align}
with the initial condition (here and in the following we take all interaction coefficients with at least one negative index to be identically zero)
\begin{equation}
\chi_{0000}=\frac{100}{77} \, .
\end{equation} 

\subsection{Recurrence relation for the $G_{mnpq}$ integrals}
To get the recurrence for the $G_{mnpq}$ integral (\ref{Gijkl}), in the auxiliary integral
\[
\int_{0}^{\frac{\pi}{2}}\text{d}x\, \frac{\mu(x)}{sin^2 x}\mu(x)\nu'(x)e_{i}(x)e_{j}(x)e_{k}(x)e_{l}(x)
\]
we either (1) use the identity (\ref{identity1}) for $\mu \nu' e_m$, or (2) use the identity
\begin{equation}
\label{eq:mu_nu_prim}
\mu \nu' = 2 - d - 2 \sin^2 x \, . 
\end{equation}
Equating the results of these two operations we get (for $d=4$)
\[
 A_{-}(m) G_{mnpq} + B(m) G_{(m+1)npq} + C(m) G_{(m-1)npq} = -2 G_{mnpq} -2 \chi_{mnpq} \, .
\]
Thus the recurrence relation for the $G_{mnpq}$ integral (totally symmetric in its indices) reads
\begin{align}
G_{mnpq}  = &\frac{1}{\sqrt{m (m+5)}} \left[-2 (2m + 5) \chi_{(m-1)npq} + \frac{m^2 + 4m + 5}{2 m+3} 4 G_{(m-1)npq} \right.
\nonumber\\
& \hskip 24mm \left.- \frac{2m + 5}{2m + 3} \sqrt{(m-1) (m+4)} G_{(m-2)npq} \right]
\end{align}
with the initial condition
\begin{equation}
G_{0000}=\frac{240}{77} \, .
\end{equation} 

\subsection{Recurrence relation for the $K_{mnp}$ integrals}
To find the recurrence relations for the $K_{mnp}$ integrals (\ref{Kijk}) we combine the methods of two previous subsections. First, in an auxiliary integral 
\[
\int_{0}^{\frac{\pi}{2}}\text{d}x\, \frac{\mu}{\sin^2 x} \mu \nu' e_{m}(x)e_{n}(x)e_{p}(x)
\]
we either (1) use the identity (\ref{identity1}) for $\mu \nu' e_m$, or (2) use the identity (\ref{eq:mu_nu_prim}) to express the $K_{mnp}$ integral in terms of  an integral $\sigma_{mnp}$ totally symmetric in its indices: 
\begin{equation}
\label{eq:SIGMA_mnp}
\sigma_{mnp} = \int_{0}^{\frac{\pi}{2}}\text{d}x\, \mu e_{m}(x)e_{n}(x)e_{p}(x) \, .
\end{equation}
Equating the results of these two operations we get:
\[
 A_{-}(m) K_{mnp} + B(m) K_{(m+1)np} + C(m) K_{(m-1)np} = -2 K_{mnp} -2 \sigma_{mnp} \, .
\]
Now, to get the recurrence relation for the $\sigma_{mnp}$ integral we consider another auxiliary integral 
\begin{equation}
\label{tildeSIGMAmnp}
\tilde \sigma_{mnp} = \int_{0}^{\frac{\pi}{2}} \text{d}x\, \mu^2(x) \nu'(x) e_{m}(x) e_{n}(x) e_{p}(x) \, .
\end{equation}
In this integral we either (1) use the identity (\ref{identity1}) for $\mu \nu' e_m$, or (2) integrate by parts using $\mu' \nu = d-1$, the identity (\ref{identity2}), and the definition (\ref{eq:SIGMA_mnp}). Equating the results of these two operations we get (for $d=4$): 
\begin{align}
  & A_{-}(m) \sigma_{mnp} + B(m) \sigma_{(m+1)np} + C(m) \sigma_{(m-1)np} 
\nonumber\\
= & - \frac{1}{2} A_{+}(m) \sigma_{mnp} - \frac{\omega_m}{2} B(m) \sigma_{(m+1)np} + \frac{\omega_m}{2} C(m) \sigma_{(m+1)np} 
\nonumber\\
  & - \frac{1}{2} A_{+}(n) \sigma_{mnp} - \frac{\omega_n}{2} B(n) \sigma_{m(n+1)p} + \frac{\omega_n}{2} C(n) \sigma_{m(n-1)p} 
\nonumber\\
  & - \frac{1}{2} A_{+}(p) \sigma_{mnp} - \frac{\omega_p}{2} B(p) \sigma_{mn(p+1)} + \frac{\omega_p}{2} C(p) \sigma_{mn(p-1)} - 6  \sigma_{mnp}
\label{SIGMAmnp_aux}
\end{align}
To eliminate $\sigma_{(n+1)pm}$ and $\sigma_{(p+1)mn}$ from the equation above, we use the identity (\ref{identity1}) in sequence for $\mu \nu' e_m$, $\mu \nu' e_n$, $\mu \nu' e_p$ to get from (\ref{tildeSIGMAmnp}) a sequence of identities 
\begin{align}
  & A_{-}(m) \sigma_{mnp} + B(m) \sigma_{(m+1)np} + C(m) \sigma_{(m-1)np}  
\nonumber\\
= & A_{-}(n) \sigma_{npm} + B(n) \sigma_{(n+1)pm} + C(n) \sigma_{(n-1)pm}  
\nonumber\\
= & A_{-}(p) \sigma_{pmn} + B(p) \sigma_{(p+1)mn} + C(p) \sigma_{(p-1)mn}  
\end{align}  
that can be solved for $\sigma_{(n+1)pm}$ and $\sigma_{(p+1)mn}$. Finally we get
\begin{align}
K_{mnp}  = &\frac{1}{\sqrt{m (m+5)}} \left[-2 (2m + 5) \sigma_{(m-1)np} + \frac{m^2 + 4m + 5}{2 m+3} 4 K_{(m-1)np} \right.
\nonumber\\
& \hskip 23mm \left.- \frac{2m + 5}{2m + 3} \sqrt{(m-1) (m+4)} K_{(m-2)np} \right]
\end{align}
and
\begin{align}
\sigma_{mnp} = & \frac{1}{(9+m+n+p+q) \sqrt{m (m+5)}} 
\nonumber\\
\times & \left\{ \frac{2 \sigma_{(m-1)np}}{(2m + 3) (2n + 5) (2p + 5)}
\left[5 \left( 100 + 60(n+p) + 5 \left( n^2 + p^2 \right) \right.\right.\right.
\nonumber\\
&  \left. \left. + 32 np + 2 \left(n^2 p + p^2n \right) \right) + m (m+4) ( 25 + 20(n+p) + 12np) \right]
\nonumber\\
& + \left[ \frac{m-n-p-5}{2 m+3} \sqrt{(m-1) (m+4)} \sigma_{(m-2)np} + \frac{2 (n+3)}{2 n+5} \sqrt{n (n+5)} \sigma_{(m-1)(n-1)p} \right.
\nonumber\\
& \left. \left. + \frac{2 (p+3)}{2 p+5} \sqrt{p (p+5)} \sigma_{(m-1)n(p-1)} \right] (2m + 5) \right\} 
\end{align}
with the initial conditions
\begin{equation}
K_{000}=\frac{3\sqrt{30}}{7} \quad \mbox{and} \quad \sigma_{000} = \frac{4\sqrt{10}}{7 \sqrt{3}} \, .
\end{equation} 

\subsection{Recurrence relations for the $Y_{mnpq}$ integrals}
Using the identity (\ref{identity2}) for $\mu \nu e_m'$ in the definition of the $Y_{mnpq}$ integral (\ref{Yijkl}), $Y_{mnpq}$ can be given in terms of integrals $\gamma_{mnpq}$, symmetric in the first and the second pairs of indices:
\begin{equation}
\label{GAMMAmnpq}
\gamma_{mnpq} = \int_{0}^{\frac{\pi}{2}} \text{d}x\, \mu(x) e_{m}(x) e_{n}(x) e_{p}'(x) e_{q}'(x) \, ,
\end{equation}
namely
\begin{equation}
\label{Ymnpq_in_GAMMA}
Y_{mnpq} = \frac{1}{2} A_{+}(m) \gamma_{mnpq} + \frac{\omega_m}{2} B(m) \gamma_{(m+1)npq} - \frac{\omega_m}{2} C(m) \gamma_{(m-1)npq} \, .
\end{equation}
Now, to get the recurrence relations in the first pair of (symmetric) indices for the $\gamma_{mnpq}$ integral we consider another auxiliary integral 
\begin{equation}
\label{tildeGAMMAmnpq}
\tilde \gamma_{mnpq} = \int_{0}^{\frac{\pi}{2}} \text{d}x\, \mu^2(x) \nu'(x) e_{m}(x) e_{n}(x) e_{p}'(x) e_{q}'(x) \, .
\end{equation}
In this integral we use identity (\ref{identity1}) either for  $\mu \nu' e_m$ or $\mu \nu' e_n$ to get
\begin{align}
  & A_{-}(m) \gamma_{mnpq} + B(m) \gamma_{(m+1)npq} + C(m) \gamma_{(m-1)npq} 
\nonumber\\
= & A_{-}(n) \gamma_{mnpq} + B(n) \gamma_{m(n+1)pq} + C(n) \gamma_{m(n-1)pq}
\label{GAMMAmnpq_aux}
\end{align}
Then, integrating 
\[
Y_{mnpq}+Y_{nmpq} = \int_0^{\pi/2} dx \, \nu \left( e_m e_n \right)' \left( \mu e_p' \right) \left( \mu e_q' \right)
\] 
by parts and using (\ref{identity1}) for $\mu \nu' e_m$, and the eigen equation for $\left( \mu e_p'\right)'$ and $\left( \mu e_p'\right)'$, we get:
\begin{align}
\label{Ysymmetrized}
Y_{mnpq}+Y_{nmpq} & =  - A_{-}(m) \gamma_{mnpq} - B(m) \gamma_{(m+1)npq} - C(m) \gamma_{(m-1)npq}  
\nonumber\\
& + \omega_p^2 X_{qpmn} + \omega_q^2 X_{pqmn} - 8 \lambda_{qpmn} - 8 \lambda_{pqmn} \, ,
\end{align}
where 
\begin{equation}
\lambda_{qpmn} = \int_0^{\pi/2} dx \, \frac{\mu^2 \nu}{\sin^2 x} e_q' e_p e_m e_n  
\end{equation}
can be easily expressed with a use of (\ref{identity2}) in terms of $G_{qpmn}$ integrals:
\begin{equation}
\label{eq:lambdaRec}
\lambda_{qpmn} = \frac{1}{2} A_{+}(q) G_{qpmn} + \frac{\omega_q}{2} B(q) G_{(q+1)pmn} - \frac{\omega_q}{2} C(q) G_{(q-1)pmn} \, .
\end{equation} 
Now, equations (\ref{Ymnpq_in_GAMMA}), (\ref{GAMMAmnpq_aux}), (\ref{Ysymmetrized}) and (\ref{eq:lambdaRec}) can be solved to yield the recurrence relation in the first pair of indices of the $\gamma_{mnpq}$ integral. In particular, setting $p=q=0$, we get 
\begin{align}
& \gamma_{mn00} = \frac{1}{(6+m+n) \sqrt{n(n+5)}} \left( \left( \frac{(2n+5)(24m+55)}{2(2m+5)} + \frac{5(2m+7)}{2(2n+3)} \right) \gamma_{m(n-1)00} \right.
\nonumber\\
& + (2n+5) \left( \frac{2(m+3)}{2m+5} \sqrt{m(m+5)} \gamma_{(m-1)(n-1)00} + \frac{n-m-2}{2n+3}\sqrt{(n-1)(n+4)} \gamma_{m(n-2)00} \right.
\nonumber\\
& \left. \left. + 72 X_{00m(n-1)} + \frac{16}{7} \left( 10 G_{00m(n-1)} - 3 \sqrt{6} G_{01m(n-1)} \right) \right) \right) \, .
\label{GAMMAmnpq_1}
\end{align}
To get the recurrence relation in the second pair of (symmetric) indices for the $\gamma_{mnpq}$ integral we consider again the auxiliary integral (\ref{tildeGAMMAmnpq}) where we either (1) use the identity (\ref{identity1}) for $\mu\nu'e_n$ or (2) use the identity (\ref{identity3}) for $\mu\nu'e_q'$. Equating the results of these two operations we get
\begin{align}
& A_{-}(n) \gamma_{mnpq} + B(n) \gamma_{m(n+1)pq} + C(n) \gamma_{m(n-1)pq} =  -\frac{16 - 4\omega_q^2 + 3 \omega_q^4}{(\omega_q^2 - 4)(\omega_q^2 - 1)} \gamma_{mnpq}
\nonumber\\
& + B(q) \frac{\omega_q}{\omega_{q+1}} \gamma_{mnp(q+1)} + C(q) \frac{\omega_q}{\omega_{q-1}} \gamma_{mnp(q-1)} + \frac{32}{\omega_q^2 - 4} \frac{\mu\nu}{\sin^2 x} \lambda_{pqmn}
\end{align}
Solving for $\gamma_{mnp(q+1)}$ and shifting the index $q+1 \rightarrow q$ we finally get
\begin{align}
& \gamma_{mnpq} = \frac{1}{\sqrt{q(q+5)}} \left( \left( 
\frac{1}{(2n + 5)(2n + 7)} \left( \frac{2 (15 + 12n + 2n^2)}{q + 2} - 5 (7 + 2q) \right) \right. \right. 
\nonumber\\
& \left.  \hskip 36mm
+ \frac{3 (3q + 7)}{(q + 1)(2q + 3)} \right) \gamma_{mnp(q-1)}
\nonumber\\
& + (q + 3)(2q + 5) \left( 
 \frac{\sqrt{(n+1)(n + 6)}}{(q + 2)(2n + 7)} \gamma_{m(n+1)p(q-1)}
- \frac{\sqrt{(q-1)(q + 4)}}{(q + 1)(2q + 3)} \gamma_{mnp(q-2)} \right.
\nonumber\\
& \left. \hskip 30mm
+ \frac{\sqrt{n(n + 5)}}{(q + 2)(2n + 5)} \gamma_{m(n-1)p(q-1)} \right)
\nonumber\\
& \frac{(2q + 5)}{(q + 1)(q + 2)} \left( 
  \frac{8(p + 3) \sqrt{p(p + 5)}}{2p + 5} G_{mn(p-1)(q-1)} 
+ \frac{16 (25 + 18 p + 3 p^2)}{(2p+5)(2p+7)} G_{mnp(q-1)} \right.
\nonumber\\
& \left. \left. \hskip 25mm
- \frac{8(p + 3) \sqrt{(p+1)(p + 6)}}{2p + 7} G_{mn(p+1)(q-1)}  \right) \right) \, .
\label{GAMMAmnpq_2}
\end{align}
Equations (\ref{GAMMAmnpq_1}) and (\ref{GAMMAmnpq_2}), together with the initial condition
\begin{equation}
\gamma_{0000}=\frac{80}{11} 
\end{equation} 
provide the complete set of recurrence relations for the $\gamma_{mnpq}$ integrals.
\subsection{Recurrence relations for the $W_{ijkk}$ integrals}
To find the recurrence relations for the $W_{ijkk}$ integrals (\ref{Wijkl}) we consider more general $W_{ijkl}$ integrals, 
\begin{equation}
W_{ijkl} =\int_{0}^{\frac{\pi}{2}}\text{d}x\,e_{i}(x)e_{j}(x)\mu(x)\nu(x)\int_{0}^{x}\text{d}y\, e_{k}(y) e_{l}(y) \mu(y) \, ,
\label{eq:Wijkl}
\end{equation}
and in the auxiliary integral 
\[
\int_{0}^{\frac{\pi}{2}}\text{d}x\,e_{i}(x)e_{j}(x)\mu(x)\nu(x)\int_{0}^{x}\text{d}y\, e_{k}(y) e_{l}(y) \mu(y) \mu(y) \nu'(y) 
\]
we use the identity (\ref{identity1}) in sequence for $\mu \nu' e_{k}$ and $\mu \nu' e_{l}$ and then substitute $l=k+1$ to get
\begin{align}
  & A_{-}(k) W_{ijk(k+1)} + B(k) W_{ij(k+1)(k+1)} + C(k) W_{ij(k-1)(k+1)} 
\nonumber\\
= & A_{-}(k+1) W_{ijk(k+1)} + B(k+1) W_{ijk(k+2)} + C(k+1) W_{ijkk} \, .
\label{eq:eq_W}
\end{align}
Then we use identities~
\footnote{They are particular cases of a general identity
\[
(\omega_k^2 - \omega_l^2) W_{ijkl} = X_{lijk} - X_{kijl} 
\]
that is easy to establish using eigen equation (\ref{eigenEq}).}
\begin{align}
W_{ijk(k+1)} &= - \frac{1}{4 (\omega_k + 1)} \left( X_{(k+1)ijk} - X_{kij(k+1)}\right)
\nonumber\\
W_{ijk(k+2)} &= - \frac{1}{8 (\omega_k + 2)} \left( X_{(k+2)ijk} - X_{kij(k+2)}\right)
\nonumber\\
W_{ij(k-1)(k+1)} &= - \frac{1}{8 \omega_k} \left( X_{(k+1)ij(k-1)} - X_{(k-1)ij(k+1)}\right) \, 
\nonumber
\end{align}
to solve (\ref{eq:eq_W}) for $W_{ij(k+1)(k+1)}$. Finally, shifting the index $k+1 \rightarrow k$, we get
\begin{align}
W_{ijkk} &= W_{ij(k-1)(k-1)} 
\nonumber\\
& - \frac{5}{(\omega_k - 3)(\omega_k^2 - 1)} \frac{1}{\sqrt{k(k+5)}} \left( X_{kij(k-1)} - X_{(k-1)ijk}\right) 
\nonumber\\
& - \frac{\omega_k - 1}{8 \omega_k (\omega_k + 1)} \sqrt{\frac{(k+1)(k+6)}{k(k+5)}} \left( X_{(k+1)ij(k-1)} - X_{(k-1)ij(k+1)}\right)
\nonumber\\
& + \frac{\omega_k - 1}{8 (\omega_k-3) (\omega_k - 2)} \sqrt{\frac{(k-1)(k+4)}{k(k+5)}} \left( X_{kij(k-2)} - X_{(k-2)ijk}\right) \, .
\end{align}
Thus the $W_{ijkk}$ integrals are given in terms of $W_{ij00}$ integrals and $X_{abcd}$ integrals. Now, to find the recurrence for the  $W_{ij00}$ integrals we consider auxiliary integrals
\begin{align}
  & 2 \int_0^{\pi/2} dx\, \mu(x) \nu(x) \mu(x) \nu'(x) e_i(x) e_j(x) \int_0^x dy \mu(y) e_k(y) e_l(y) 
\nonumber\\
+ &  \int_0^{\pi/2} dx\, \mu(x) \nu(x) e_i(x) e_j(x) \int_0^x dy \mu(y) \mu(y) \nu'(y)e_k(y) e_l(y)
\end{align}
and we either (1) use the identity (\ref{identity1}) for $\mu \nu' e_i$ and  $\mu \nu' e_k$, or (2) integrate by parts using $\mu' \nu = d-1$, the identity (\ref{identity2}), and the definition (\ref{eq:Wijkl}). Then, in the second of the auxiliary integrals we either (3) use the identity (\ref{identity1}) for $\mu \nu' e_i$, or (4) use the identity (\ref{identity1}) for $\mu \nu' e_j$. These two pairs of operations result in the system of two equations that can be solved for $W_{(i+1)jkl}$ and $W_{i(j+1)kl}$. Then shifting the index $i+1 \rightarrow i$ and setting $k=l=0$ we get:
\begin{align}
W_{ij00}=& \frac{1}{(i+j+7) \sqrt{j(j+5)}} \left( \left( \frac{(12 i + 25)(2 j + 5)}{2(2i+5)} + \frac{5(2i+9)}{2(2j+3)}\right) W_{i(j-1)00} \right.
\nonumber\\
& + (2j + 5) \left( \frac{2(i+3)}{2i+5}\sqrt{i(i+5)}W_{(i-1)(j-1)00} + \frac{j-i-3}{2j+3}\sqrt{(j-1)(j+4)}W_{i(j-2)00} \right.
\nonumber\\
& + \left. \left. \frac{\sqrt{3}}{14 \sqrt{2}} \left( X_{10i(j-1)} - X_{01i(j-1)}\right) \right) \right) \,,
\end{align}  
with the initial condition
\[
W_{0000} = \frac{358}{3003} \,.
\]

\subsection{Recurrence relations for the $\bar{W}_{ijkk}$ integrals}
The recurrence relations for the $\bar{W}_{ijkk}$ integrals (\ref{WSijkk}) can be obtained in close analogy to the case of $W_{ijkk}$ integrals described in the previous subsection. To find the recurrence relations for the $\bar{W}_{ijkk}$ integrals we consider more general $\bar{W}_{ijkl}$ integrals, 
\begin{equation}
\bar{W}_{ijkl} =\int_{0}^{\frac{\pi}{2}}\text{d}x\,e'_{i}(x)e'_{j}(x)\mu(x)\nu(x)\int_{0}^{x}\text{d}y\, e_{k}(y) e_{l}(y) \mu(y) \, ,
\label{eq:WSijkl}
\end{equation}
and in the auxiliary integral 
\[
\int_{0}^{\frac{\pi}{2}}\text{d}x\,e_{i}'(x)e_{j}'(x)\mu(x)\nu(x)\int_{0}^{x}\text{d}y\, e_{k}(y) e_{l}(y) \mu(y) \mu(y) \nu'(y) 
\]
we use the identity (\ref{identity1}) in sequence for $\mu \nu' e_{k}$ and $\mu \nu' e_{l}$ and then substitute $l=k+1$ to get
\begin{align}
  & A_{-}(k) \bar{W}_{ijk(k+1)} + B(k) \bar{W}_{ij(k+1)(k+1)} + C(k) \bar{W}_{ij(k-1)(k+1)} 
\nonumber\\
= & A_{-}(k+1) \bar{W}_{ijk(k+1)} + B(k+1) \bar{W}_{ijk(k+2)} + C(k+1) \bar{W}_{ijkk} \, .
\label{eq:eq_WS}
\end{align}
Then we use identities~
\footnote{They are particular cases of a general identity
\[
(\omega_k^2 - \omega_l^2) \bar{W}_{ijkl} = Y_{lkij} - Y_{klij} 
\]
that is easy to establish using eigen equation (\ref{eigenEq}).}
\begin{align}
\bar{W}_{ijk(k+1)} &= - \frac{1}{4 (\omega_k + 1)} \left( Y_{(k+1)kij} - Y_{k(k+1)ij}\right)
\nonumber\\
\bar{W}_{ijk(k+2)} &= - \frac{1}{8 (\omega_k + 2)} \left( Y_{(k+2)kij} - Y_{k(k+2)ij}\right)
\nonumber\\
\bar{W}_{ij(k-1)(k+1)} &= - \frac{1}{8 \omega_k} \left( Y_{(k+1)(k-1)ij} - Y_{(k-1)(k+1)ij}\right) \, 
\nonumber
\end{align}
to solve (\ref{eq:eq_WS}) for $\bar{W}_{ij(k+1)(k+1)}$. Finally, shifting the index $k+1 \rightarrow k$, we get
\begin{align}
\bar{W}_{ijkk} &= \bar{W}_{ij(k-1)(k-1)} 
\nonumber\\
& - \frac{5}{(2k+3)(2k+5)(2k+7)} \frac{1}{\sqrt{k(k+5)}} \left( Y_{k(k-1)ij} - Y_{(k-1)kij}\right) 
\nonumber\\
& - \frac{2k+5}{16 (k+3)(2k+7)} \sqrt{\frac{(k+1)(k+6)}{k(k+5)}} \left( Y_{(k+1)(k-1)ij} - Y_{(k-1)(k+1)ij}\right)
\nonumber\\
& + \frac{2k+5}{16 (k+2) (2k+3)} \sqrt{\frac{(k-1)(k+4)}{k(k+5)}} \left( Y_{k(k-2)ij} - Y_{(k-2)kij}\right) \, .
\end{align}
Thus the $\bar{W}_{ijkk}$ integrals are given in terms of $\bar{W}_{ij00}$ integrals and $Y_{abcd}$ integrals. Now, to find the recurrence for the  $\bar{W}_{ij00}$ integrals we integrate (\ref{eq:WSijkl}) by parts and (using eigenequation (\ref{eigenEq}) for $\left(\mu e_i'\right)'$) we get
\[
\bar{W}_{ijkl} = \omega_i^2 W_{ijkl} - 8 T_{ijkl} - N_{ijkl} - X_{ijkl} \, ,
\]
where
\begin{equation}
T_{ijkl} = \int_0^{\pi/2} dx \, \frac{\mu(x)\nu(x)}{\sin^2 x} e_i(x) e_j(x) \int_0^{x} dy \,  e_{k}(y) e_{l}(y) \mu(y)
\label{eq:Tijkl}
\end{equation}
and
\[
N_{ijkl} = \int_0^{\pi/2} dx \, \mu(x)\nu'(x) e_i'(x) e_j(x) \int_0^{x} dy \,  e_{k}(y) e_{l}(y) \mu(y)
\]
Since the left hand side of (\ref{eq:WSijkl}) is symmetric in $ij$ indices we can write
\[
2 \bar{W}_{ijkl} = \omega_i^2 W_{ijkl} + \omega_j^2 W_{ijkl} - 16 T_{ijkl} - N_{ijkl} - N_{jikl} - X_{ijkl} - X_{jikl} \, ,
\]
Now, integrating twice by parts and using $\left(\mu \nu'\right)' = -4 \mu \nu$ and $\mu'\nu=d-1$, it can be easy established that
\begin{align*}
N_{ijkl} + N_{jikl} & = 4 W_{ijkl} - \int_{0}^{\pi/2} \mu^2 \nu' e_i e_j e_k e_l 
\\
& = 4 W_{ijkl} + 2(d-1) \chi_{ijkl} + X_{ijkl} + X_{jkli} + X_{klij} + X_{lijk}
\end{align*}
thus we finally get (for $d=4$):
\begin{align}
\bar{W}_{ij00}=& 2(17+i(i+6)+j(j+6)) W_{ij00} - 8 T_{ij00} - X_{00ij} - X_{ij00} - X_{ji00} - 3\chi_{ij00}
\label{eq:RecWSij00}
\end{align}  
with the initial condition
\[
\bar{W}_{0000} = \frac{1216}{1001} \,.
\]
To find the recurrence relations for the $T_{ijkl}$ integrals (\ref{eq:Tijkl}), needed in (\ref{eq:RecWSij00}), we consider an auxiliary integral 
\[
\int_{0}^{\frac{\pi}{2}} dx \, \frac{\mu(x)\nu(x)}{\sin^2 x} \mu \nu' e_{i}(x)e_{j}(x) \int_0^{x} dy \,  e_{k}(y) e_{l}(y) \mu(y)
\]
and we either (1) use the identity (\ref{identity1}) for $\mu \nu' e_j$, or (2) use the identity (\ref{eq:mu_nu_prim}). This yields
\[
 A_{-}(j) T_{ijkl} + B(j) T_{i(j+1)kl} + C(j) T_{i(j-1)kl} = (2-d) T_{ijkl} -2 W_{ijkl} \, 
\]
and it finally gives
\begin{align}
T_{ij00} =& \frac{1}{\sqrt{j(j+5)}} \left( - 2 (2 j + 5) W_{i(j-1)00} + \frac{4(2+(j+1)(j+3))}{2j+3} T_{i(j-1)00} \right. 
\nonumber\\
& \left. - \frac{(2j+5)\sqrt{(j-1)(j+4)}}{2j+3} T_{i(j-2)00}\right) \, ,
\end{align}
with the initial condition
\[
T_{0000} = \frac{8}{33} \, .
\]

\subsection{Recurrence relation and closed form expressions for the $V_{mn}$ integrals}
To find the recurrence for the $V_{mn}$ integrals (\ref{Vij}) (symmetric in their indices), we consider an auxiliary integral
\[
2 \int_0^{\pi/2} dx\, \mu \nu \mu \nu' e_m e_n
\]
and we either (1) use the identity (\ref{identity1}) for $\mu \nu' e_m$, or (2) use the identity (\ref{identity1}) for $\mu \nu' e_n$ or (3) integrate by parts using $\mu' \nu = d-1$ and the identity (\ref{identity2}) for $\mu \nu e_m'$ and $\mu \nu e_n'$. Equating the results of these three operations we get (for $d=1$):
\begin{align}
  & 2 \left( A_{-}(m) V_{mn} + B(m) V_{(m+1)n} + C(m) V_{(m-1)n} \right) 
\nonumber\\
= & 2 \left( A_{-}(n) V_{mn} + B(n) V_{m(n+1)} + C(n) V_{m(n-1)} \right) 
\nonumber\\
=  & -6 V_{mn} - \frac{1}{2} A_{+}(m) V_{mn} - \frac{\omega_m}{2} B(m) V_{(m+1)n} + \frac{\omega_m}{2} C(m) V_{(m-1)n} 
\nonumber\\
  & \hskip 14mm - \frac{1}{2} A_{+}(n) V_{mn} - \frac{\omega_n}{2} B(n) V_{m(n+1)} + \frac{\omega_n}{2} C(n) V_{m(n-1)}
\end{align}
These system can be solved for $V_{(m+1)n}$ and $V_{m(n+1)}$. Shifting the index $m+1 \rightarrow m$ we finally get
\begin{align}
V_{mn}=& \frac{1}{(m+n+7) \sqrt{m(m+5)}} \left(\frac{2 \left( m (m+4) (12n+25) + 5 \left( n^2+16n+30\right) \right)} {(2m+3)(2n+5)} V_{(m-1)n} \right.
\nonumber\\
& \left. + (2m + 5) \left( \frac{m-n-3}{2m+3}\sqrt{(m-1)(m+4)} V_{(m-2)n} + \frac{2(n+3)}{2n+5}\sqrt{n(n+5)} V_{(m-1)(n-1)}  \right) \right)\,,
\label{VmnRec}
\end{align}  
with the initial condition
\begin{equation}
V_{00} = \frac{4}{7} \,.
\label{V00}
\end{equation}
Interestingly, the solution of the recurrence relations (\ref{VmnRec}, \ref{V00}) can be found in a closed form~
\footnote{The solutions can be found by \textit{Mathematica} if sufficient number of initial values of the sequences $V_{mm}$, $V_{(m+1)m}$, $V_{(m+2)m}$, $\dots$ are generated.}:
\begin{align}
V_{mm} &= \frac{2 (m+1) (m+2) (4m + 15)}{3 (2m + 5) (2 m + 7)} \,,
\nonumber\\
V_{m(m-1)} &= \frac{(m+1) (8m + 25)}{6 (2m + 5)} \sqrt{\frac{m}{m+5}} \,,
\nonumber\\
V_{mn} &\stackrel{m-n>1}{=} \frac{2}{3} (n+3) \sqrt{\frac{(n+1)^{\overline{5}}}{(m+1)^{\overline{5}}}} \,,
\label{VmnSol}
\end{align}
where $n^{\overline{k}} := n(n+1)...(n+k-1)$, $k>0$. 

\subsection{Recurrence relation and closed form expressions for the $A_{mn}$ integrals}
To find the recurrence for the $A_{mn}$ integrals (\ref{Aij}) (symmetric in their indices), we integrate by parts using
\[
\mu \nu e_m' e_n' = \left( \mu \nu e_m e_n'\right)' - \nu e_m \left( \mu e_n' \right)' - \mu \nu' e_m e_n'
\]
and the eigen equation
\[
\left( \mu e_n' \right)' = - \mu \omega_n^2 e_n + \frac{8 \mu}{\sin^2 x} e_n
\]
Then symmetrizing the result as $A_{mn} = \left( A_{mn} + A_{nm} \right)/2$ and using
\[
-\frac{1}{2} \mu \nu' \left( e_m e_n' + e_m' e_n\right) = -\frac{1}{2} \left( \mu \nu' e_m e_n \right)' + \frac{1}{2} \left( \mu \nu' \right)'  e_m e_n
\]
together with $ \left( \mu \nu' \right)' = -4 \mu \nu$ we finally get
\begin{equation}
A_{mn} = \frac{1}{2} \left( \omega_m^2 + \omega_n^2 - 4 \right) V_{mn} - 8 Q_{mn} \, ,
\label{AmnRec}
\end{equation}
with 
\begin{equation}
Q_{mn} = \int_{0}^{\frac{\pi}{2}}\text{d}x\, \frac{\mu}{\sin^2 x} e_m e_n \, .
\label{Qmn}
\end{equation}
The recurrence relation for the $Q_{mn}$ integrals (symmetric in their indices) can be easily obtained in analogy to the $K_{mnp}$ integrals: in an auxiliary integral 
\[
\int_{0}^{\frac{\pi}{2}}\text{d}x\, \frac{\mu}{\sin^2 x} \mu \nu' e_m e_n 
\] 
we either (1) use the identity (\ref{identity1}) for $\mu \nu' e_m$, or (2) use the identity (\ref{eq:mu_nu_prim}). Equating the results of these two operations we get (for $d=4$):
\begin{equation}
 A_{-}(m) Q_{mn} + B(m) Q_{(m+1)n} + C(m) Q_{(m-1)n} = -2 Q_{mn} - 2 V_{mn} \, .
\end{equation}
Shifting the index $m+1 \rightarrow m$ we finally get
\begin{equation}
Q_{mn} = \frac{2m + 5}{\sqrt{m(m+5)}} \left( \frac{4 \left( m^2 + 4m + 5 \right)}{(2m+5)(2m+3)} Q_{(m-1)n} - \frac{\sqrt{(m-1)(m+4)}}{2m+3} Q_{(m-2)n} - 2 V_{(m-1)n}\right) \,,
\label{QmnRec}
\end{equation}  
with the initial condition
\begin{equation}
Q_{00} = 2 \,.
\label{Q00}
\end{equation}
Interestingly, the solution of the recurrence relations (\ref{AmnRec}, \ref{QmnRec}, \ref{Q00}) can be found in a closed form~
\footnote{The solutions can be found by \textit{Mathematica} if sufficient number of initial values of the sequences $A_{mm}$, $A_{(m+1)m}$, $A_{(m+2)m}$, $\dots$ are generated.}:
\begin{align}
A_{mm} &= \frac{4 (m+1) (m+2) (m+3) (4m^2 + 18m + 15)}{3 (2m + 5) (2 m + 7)} \,,
\nonumber\\
A_{m(m-1)} &= \frac{2 (m+1) (m+2) (4m^2 + 11m + 5)}{3 (2m + 5)} \sqrt{\frac{m}{m+5}} \,,
\nonumber\\
A_{mn} &\stackrel{m-n>1}{=} \frac{4}{3} (n+3) (15 + 6(2n-m) + 2n^2 - m^2) \sqrt{\frac{(n+1)^{\overline{5}}}{(m+1)^{\overline{5}}}} \,.
\label{AmnSol}
\end{align}

\section{Preliminary numerical results}
\label{PreliminaryNumericalResults}
As it was stressed in Sec.~\ref{Introduction}, investigating the problem of the AdS stability by solving numerically the Einstein equations (\ref{Einsteineq}), we can never have access to the $\varepsilon \rightarrow 0$ limit (as the instability can be expected to be revealed at the $\mathcal{O}\left(\varepsilon^{-2}\right)$ time-scale at the earliest). On the other hand due the the scaling symmetry 
\begin{equation}
C_l(\tau) \rightarrow \varepsilon \, C_l\left(\varepsilon^2 \tau \right) \quad \mbox{and} \quad
\Phi_l(\tau) \rightarrow \Phi_l\left(\varepsilon^2 \tau \right) \, ,  
\end{equation}
i.e. if $C_l(\tau)$ and $\Phi_l(\tau)$ are a solutions to (\ref{ResonantSystem}) so are $\varepsilon \, C_l\left(\varepsilon^2 \tau \right)$ and $\Phi_l\left(\varepsilon^2 \tau \right)$. Thus,  the solutions of the resonant system (\ref{ResonantSystem}) capture the dynamics at $\mathcal{O}\left(\varepsilon^{-2}\right)$ time-scale exactly under assumption of neglecting the effects of non-resonant terms (the neglected higher order terms affect the dynamics on longer time-scales $\mathcal{O}\left(\varepsilon^{-k}\right)$ with integer $k>2$). Of course, to solve (\ref{ResonantSystem}) numerically one has to introduce some truncation in the number of modes present in the system, i.e. to introduce upper limit $N$ in the sums in (\ref{ResonantSystem}). Anyway, it would be desirable to solve the resonant system (\ref{ResonantSystem}) numerically for some model initial data (for example two-modes initial data that were already intensively studied in the past for the massless scalar field in $3+1$ \cite{bbgll_PRL113, br_PRL115, df_JHEP1512}, in $4+1$ \cite{bmr_PRL115}, and in higher dimensions \cite{d_1606.02712}) to check for a convergence between 
\begin{enumerate}
\item
the solutions of (\ref{Einsteineq}) with initial data $B(0,x) = \varepsilon f(x)$ and $\dot B(0,x) = \varepsilon g(x)$ in the $\varepsilon \rightarrow 0$ limit,
\item
the solutions of (\ref{ResonantSystem}) with initial data inferred from $B_1(0,x) = f(x)$ and $\dot B_1(0,x) = g(x)$ in the $N \rightarrow \infty$ limit,
\end{enumerate}  
and the existence of a finite-time blow-up in the resonant system, cf. \cite{bmr_PRL115}. Also, one of motivations to study higher orders in perturbation expansion \cite{r_PRD95, r_PRD96} was to lay the foundations for constructing the resonant system for arbitrarily gravitational perturbations. Although construction of such system should be conceptually straightforward after the model case \cite{cev_JHEP1410, cev_JHEP1501} and the present study, technically if would be a formidable task. Thus, before attacking such problem, it would be desirable to know if, with the presently numerically accessible cutoffs $N$, one can rely on the solutions of the resonant system (\ref{ResonantSystem}) obtained under simplifying symmetry assumptions (\ref{bcs_ansatz}). Unfortunately, it seems from the preliminary results of Maliborski \cite{m_private2} that numerical integration of the resonant system for the ansatz (\ref{bcs_ansatz}) is much more demanding then the analogous problem for the spherically symmetric massless scalar field system in $4+1$ dimensions \cite{bmr_PRL115}. Namely, even with the cutoff $N \approx 500$ it was very difficult to establish what is the decay rate of the energy power spectrum: the obtained results seemed not to converge to the decay rate $-5/3$ reported in \cite{br_APPB48}, and were giving some values between $-2$ and $-5/3$ depending on the fitting time and the range of modes used in a fit \cite{m_private2}. It would be very interesting to revisit this problem again.

\section{Acknowledgements}
We wish to thank Maciej Maliborski for his collaboration at the early stage of this project. This work was supported by the Narodowe Centrum Nauki (Poland) Grant no. 2017/26/A/ST2/530.

\appendix

\section{Vanishing of the secular terms at the second order}
\label{appA}

We prove that all the secular terms vanish at the second order in $\varepsilon$. The interaction coefficients due to quadratic nonlinearity are 
\begin{equation} \label{Kjkn}
K_{jkn} = \int^{\pi/2}_0 e_j(x) e_k(x) e_n(x) \frac{\sin x}{\cos^3 x} \, dx\; .
\end{equation}
Using $y=\cos(2x)$ and the definition of eigenfunctions (\ref{modes})
we have
\begin{equation}
e_n(y)\sim (1-y)(1+y)^2 P_n^{(3,2)}(y) \; .
\end{equation}
Then, using the formula 
\begin{equation}
P^{(\alpha, \beta)}_n(y) \sim (1-y)^{-\alpha} (1+y)^{-\beta} \frac{d^n}{d x^n}\left( (1-y)^{\alpha+n}(1+y)^{\beta+n}  \right) \, ,
\end{equation}
we get
\begin{equation}
K_{jkn} \sim \int^1_{-1} (1+y)^2 P_j^{(3,2)} P_k^{(3,2)}\frac{d^n}{d x^n}\left( (1-y)^{3+n}(1+y)^{2+n} \right) \; .
\end{equation}
Integrating by parts we find that $K_{jkn} = 0$ if 
\begin{equation} \label{K0condition}
n>j+k+2
\end{equation} (because $(1+y)^2 P_j^{(3,2)} P_k^{(3,2)}$ is the polynomial of  order $j+k+2$). For the resonant terms $\omega_n = \omega_j+\omega_k $, hence $n=3+j+k$. Thus, the coefficients of the resonant terms vanish.

\section{Calculation of $S^{(3)}_l$ and vanishing of some secular terms at the third order}
\label{appB}

To obtain $S^{(3)}_l = \scalar{S^{(3)}}{e_l}$ we follow closely the work of Craps, Evnin \& Vanhoof (2014)~\cite{cev_JHEP1410}. Our calculation is very similar to that described in Appendix A of their paper, therefore we will only give a brief picture and final results.

To get $A_2(t,x)$ from (\ref{A2integral}) in terms of the first order solution (\ref{seriesB1}) we use identities:
\begin{align}
\label{eq:iden2}
\left( \mu \left( e'_{i}e_{j} - e'_{j}e_{i} \right) \right)' &= \left( \omega_{j}^{2} -\omega_{i}^{2} \right) \mu \, e_{j}e_{i} \; ,
\\
\label{eq:iden3}
\left(\mu \left( \omega_{j}^{2} e'_{i}e_{j} - \omega_{i}^{2} e'_{j}e_{i} \right) \right)' &= \left( \omega_{j}^{2} - \omega_{i}^{2}\right) \mu \left( e'_{j}e'_{i} + \frac{8}{\sin^2 x} e_i e_j \right) \; 
\end{align}
that are easily established from the eigen equation (\ref{eigenEq}). Using these identities we get
\begin{align}
  & \frac{A_2(t,x)}{-2}
\nonumber\\
= & \nu(x) \sum_{\scriptsize{\begin{matrix} i,j\\i \neq j \end{matrix}}} \frac{\mu(x) \left[ c_i(t) c_j(t) \left( \omega_j^2 e_i'(x) e_j(x) - \omega_i^2 e_j'(x) e_i(x) \right) + \dot c_i(t) \dot c_j(t) \left( e_i'(x) e_j(x) - e_j'(x) e_i(x) \right) \right]}{ \omega_{j}^{2} - \omega_{i}^{2} }
\nonumber\\
+ & \nu(x) \sum_i \int_0^x \left[ c_i^2(t) \left( \left( e_i'(y) \right)^2 + \frac{8}{\sin^2(y)} e_i^2(y) \right) + \dot c_i^2(t) e_i^2(y) \right] \mu(y) \, dy \, .
\end{align}
Using the symmetry in $i,j$ indices under the first sum and integrating by parts and using the eigen equation (\ref{eigenEq}) under the second sum, we finally get
\begin{align}
\label{A2}
\frac{A_2(t,x)}{-2} & =  2 \nu(x) \sum_{\scriptsize{\begin{matrix} i,j\\i \neq j \end{matrix}}} \frac{ \dot c_i(t) \dot c_j(t) + \omega_j^2 c_i(t) c_j(t) }{ \omega_{j}^{2} - \omega_{i}^{2} } \mu(x) e_i'(x) e_j(x) 
\nonumber\\
& + \nu(x) \sum_i\left[ c_i^2(t) \mu(x)\, e_i'(x) e_i(x) + Q_i(t)  \int_0^x \mu(y) \, e_i^2(y) \, dy \right] \, ,
\end{align}
where $Q_i(t) = \dot c_i^2(t) + \omega_i^2 c_i^2(t)$ and $\dot Q_i \equiv 0$ from (\ref{oscilator}). Using this identity and (\ref{oscilator}) again it follows that
\begin{equation}
\label{A2dot}
\frac{\dot A_2(t,x)}{-2}  =  2 \nu(x) \sum_{i,j} c_i(t) \dot c_j(t) \mu(x) \, e_i'(x) e_j(x)  \, .
\end{equation}
Now, with (\ref{oscilator}) and (\ref{A2}, \ref{A2dot}) it is straight forward to establish that:
\begin{align}
\label{A2ddotB1el}
\frac{\scalar{A_2 \ddot B_1}{e_l}}{-2} 
& = - 2 \sum_{\scriptsize{\begin{matrix} i,j,k\\i \neq j\end{matrix}}} \frac{\omega_k^2 \, c_k}{\omega_j^2 - \omega_i^2} \left( \dot c_i \dot c_j +\omega_j^2 c_ic_j \right) \, X_{ijkl} 
- \sum_{i,k} \omega_{k}^{2} \, c_{k}\left( c_i^2 \, X_{iikl} + Q_i \, W_{klii} \right) \; ,
\\
\frac{\scalar{\dot A_2 \dot B_1}{e_l}}{-2} 
& = 2 \sum_{i,j,k} c_i \dot c_j \dot c_k \, X_{ijkl} \; ,
\\
\frac{\scalar{\frac{\displaystyle 1}{\displaystyle \sin^2 x} A_2 B_1}{e_l}}{-2} 
& = 2 \sum_{\scriptsize{\begin{matrix} i,j,k\\i \neq j\end{matrix}}} \frac{c_k}{\omega_j^2 - \omega_i^2} \left( \dot c_i \dot c_j + \omega_j^2 c_ic_j \right) \, \tilde X_{ijkl} 
+ \sum_{i,k} c_{k}\left( c_i^2 \, \tilde X_{iikl} + Q_i \, \tilde W_{klii} \right) \; ,
\end{align}
where the interaction coefficients $X_{ijkl}$, $W_{ijkl}$, $\tilde X_{ijkl}$ and $\tilde W_{ijkl}$ (i.e. integrals of products of AdS linear eigen modes and some weights) are defined in (\ref{eq:coeffs}). 

To obtain $\scalar{\delta_2 \ddot B_1}{e_l}$ and $\scalar{\dot \delta_2 \dot B_1}{e_l}$ contributions to the source $S^{(3)}_l$ we use (\ref{eq:iden2}) and integrate by parts:
\begin{align}
& \frac{\scalar{\delta_2 \ddot B_1}{e_l}}{-2} 
\nonumber\\
= &\sum_k \ddot c_k \int_0^{\pi/2} dx \, \mu(x) \, e_k(x) e_l(x) \int_0^{x} dy \, \mu(y) \nu(y) \left(B_1'^2 (t,y) + \dot B_1^2(t,y) \right)  
\nonumber\\
= & \sum_{\scriptsize{\begin{matrix} k\\k \neq l\end{matrix}}} \frac{- \omega_k^2 \, c_k}{\omega_l^2 - \omega_k^2} \underbrace{\int_0^{\pi/2} dx \, \left( \mu(x) \left( e_k'(x) e_l(x) - e_l'(x) e_k(x) \right) \right)' \int_0^{x} dy \, \mu(y) \nu(y) \left(B_1'^2 (t,y) + \dot B_1^2(t,y) \right)}_{\displaystyle \mathcal{I}_1}
\nonumber\\
- & \omega_l^2 \, c_l  \underbrace{\int_0^{\pi/2} dx \, \mu(x) \, e_l^2(x) \int_0^{x} dy \, \mu(y) \nu(y) \left(B_1'^2 (t,y) + \dot B_1^2(t,y) \right)}_{\displaystyle \mathcal{I}_2}
\end{align}
Now
\begin{align}
\mathcal{I}_1 & = - \int_0^{\pi/2} dx \, \mu(x) \left( e_k'(x) e_l(x) - e_l'(x) e_k(x) \right) \mu(x) \nu(x) \underbrace{\left(B_1'^2 (t,x) + \dot B_1^2(t,x) \right)}_{
\sum_{i,j} \left( c_i c_j e_i' e_j' + \dot c_i \dot c_j e_i e_j \right)}
\nonumber\\
& = - \sum_{i,j} \left[ \dot c_i \dot c_j \left( X_{klij} - X_{lkij} \right) + c_i c_j \left( Y_{klij} - Y_{lkij} \right) \right]
\end{align}
and
\begin{align}
\mathcal{I}_2 & = \underbrace{ \int_0^{\pi/2} dx \, \mu(x) e_l^2(x) }_{\scalar{e_l}{e_l} = 1} \int_0^{\pi/2} dx \, \mu(x) \nu(x) \left(B_1'^2 (t,x) + \dot B_1^2(t,x) \right)
\nonumber\\
& -  \int_0^{\pi/2} dx \, \mu(x) \nu(x) \left(B_1'^2 (t,x) + \dot B_1^2(t,x) \right) \int_0^{x} dy \, \mu(y) e_l^2(y) 
\nonumber\\
& = \sum_{i,j} \left( \dot c_i \dot c_j P_{ijl} + c_i c_j B_{ijl} \right) \, ,
\end{align}
where the interaction coefficients $X_{ijkl}$, $Y_{ijkl}$, $P_{ijl}$ and $B_{ijl}$ are defined in (\ref{eq:coeffs}). This gives
\begin{align}
  & \frac{\scalar{\delta_2 \ddot B_1}{e_l}}{-2} 
\nonumber\\
= & \sum_{\scriptsize{\begin{matrix} i,j,k\\k \neq l\end{matrix}}} \frac{\omega_k^2 \, c_k}{\omega_l^2 - \omega_k^2} \left[ \dot c_i \dot c_j \left( X_{klij} - X_{lkij} \right) + c_i c_j \left( Y_{klij} - Y_{lkij} \right) \right]
- \omega_l^2 \, c_l \sum_{i,j} \left( \dot c_i \dot c_j P_{ijl} + c_i c_j B_{ijl} \right) \, .
\end{align}
Similarly
\begin{align}
& \frac{\scalar{\dot \delta_2 \dot B_1}{e_l}}{-2} 
\nonumber\\
= & - \sum_{\scriptsize{\begin{matrix} i,j,k\\k \neq l\end{matrix}}} \frac{\dot c_k}{\omega_l^2 - \omega_k^2} \partial_t \left[ \dot c_i \dot c_j \left( X_{klij} - X_{lkij} \right) + c_i c_j \left( Y_{klij} - Y_{lkij} \right) \right]
+ \dot c_l \sum_{i,j} \partial_t \left( \dot c_i \dot c_j P_{ijl} + c_i c_j B_{ijl} \right) \, 
\nonumber\\
= & - \sum_{\scriptsize{\begin{matrix} i,j,k\\k \neq l\end{matrix}}} \frac{\dot c_k}{\omega_l^2 - \omega_k^2} \left\{ c_i \dot c_j \left[ -\omega_i^2 \left( X_{klij} - X_{lkij} \right) + \left( Y_{klij} - Y_{lkij} \right) \right] \right.
\nonumber\\
& \hskip 25mm 
\left. + c_j \dot c_i \left[ -\omega_j^2 \left( X_{klij} - X_{lkij} \right) + \left( Y_{klij} - Y_{lkij} \right) \right] \right\}
\nonumber\\
& + \dot c_l \sum_{i,j}  \left[ c_i \dot c_j \left( - \omega_i^2 P_{ijl} + B_{ijl}\right) + c_j \dot c_i \left( - \omega_j^2 P_{ijl} + B_{ijl}\right)\right] \, .
\end{align}
Now from
\begin{equation}
A_2' - \delta_2' = \frac{\nu'}{\nu} A_2 - \frac{16}{\sin^2 x} \mu \nu B_1^2 \; 
\end{equation}
and (\ref{A2}) we get
\begin{align}
& \scalar{ \left(A_2' - \delta_2'\right) B_1'}{e_l} 
\nonumber\\
& = -4 \sum_{\scriptsize{\begin{matrix} i,j,k\\i \neq j \end{matrix}}} \frac{ c_k \left( \dot c_i \dot c_j + \omega_j^2 c_i c_j \right) }{ \omega_j^2 -\omega_i^2} H_{ijkl} 
- 2 \sum_{i,k} c_k \left( c_i^2 H_{iikl} + Q_i M_{kli} \right) 
- 16 \sum_{i,k,j} c_i c_j c_k \tilde{X}_{klij} \; ,
\end{align}
where the interaction coefficients $H_{ijkl}$, $M_{ijk}$ and $\tilde{X}_{ijkl}$ are defined in (\ref{eq:coeffs}). Finally
\begin{equation}
\label{B1cubeel}
\scalar{\frac{1}{\sin^2 x} B_1^3}{e_l} = \sum_{i,j,k} c_i c_j c_k G_{ijkl} \, ,
\end{equation}
where the interaction coefficient $G_{ijkl}$ is defined in (\ref{Gijkl}). To make the time dependence in (\ref{A2ddotB1el}-\ref{B1cubeel}) explicit we gather some trigonometric identities (cf. (\ref{cn})):
\begin{align}
c_k c_i c_j & = \frac{1}{4} a_k a_i a_j 
\nonumber\\
& \times \left( 
  \cos\left( \theta_i - \theta_j - \theta_k \right) 
+ \cos\left( \theta_i - \theta_j + \theta_k \right) 
+ \cos\left( \theta_i + \theta_j - \theta_k \right) 
+ \cos\left( \theta_i + \theta_j + \theta_k \right) 
\right)
\\
c_k \dot c_i \dot c_j & = \frac{1}{4} a_k a_i a_j \omega_i \omega_j  
\nonumber\\
& \times \left( 
  \cos\left( \theta_i - \theta_j - \theta_k \right) 
+ \cos\left( \theta_i - \theta_j + \theta_k \right) 
- \cos\left( \theta_i + \theta_j - \theta_k \right) 
- \cos\left( \theta_i + \theta_j + \theta_k \right) 
\right)
\\
c_k \left( \dot c_i \dot c_j + \omega_j^2 c_i c_j \right)& = \frac{1}{4} a_k a_i a_j 
\nonumber\\
& \times \left[ 
  \omega_j \left( \omega_j + \omega_i \right) \left( \cos\left( \theta_i - \theta_j - \theta_k \right) + \cos\left( \theta_i - \theta_j + \theta_k \right) \right) \right.
\nonumber\\
& \hspace{2mm} \left. + \omega_j \left( \omega_j - \omega_i \right) \left( \cos\left( \theta_i + \theta_j - \theta_k \right) + \cos\left( \theta_i + \theta_j + \theta_k \right)\right) 
\right]
\end{align}
\begin{align}
c_k c_i^2 & = \frac{1}{2} a_k a_i^2 \cos(\theta_k) + \frac{1}{4} a_k a_i^2 \left( \cos \left(2 \theta_i - \theta_k \right) + \cos \left(2 \theta_i + \theta_k \right) \right) 
\\
c_k Q_i &= a_k a_i^2 \omega_i^2 \cos \theta_k
\end{align}
Using these identities it is straightforward to establish that (for the future convenience we underline some terms that are convenient to be summed up or we indicate a convenient change of indices in some other terms)
\begin{align}
  & 8 \scalar{ \frac{1}{\sin^2 x} A_2 B_1}{e_l} 
\nonumber\\
= & -8 \sum_{\scriptsize{\begin{matrix} i,j,k\\i \neq j \end{matrix}}} a_i a_j a_k \tilde{X}_{ijkl} 
\left[ 
\frac{\omega_j}{\omega_j - \omega_i} ( \cos \stackrel{i \leftrightarrow k}{\left( \theta_i - \theta_j - \theta_k \right)} + \cos \stackrel{j \leftrightarrow k}{\left( \theta_i - \theta_j + \theta_k \right)} ) \right.
\nonumber\\
& \hspace{33mm} \left. + \underline{\frac{\omega_j}{\omega_j + \omega_i} \left( \cos \left( \theta_i + \theta_j - \theta_k \right) + \cos\left( \theta_i + \theta_j + \theta_k \right) \right) }
\right]
\nonumber\\
 & -4 \sum_{i,k} a_i^2 a_k \left[ 
\left( 2 \tilde{X}_{iikl} + 4 \omega_i^2 \tilde{W}_{klii}\right) \cos(\theta_k) 
+ \underline{ \tilde{X}_{iikl} \left( \cos \left( 2 \theta_i - \theta_k \right) + \cos\left( 2 \theta_i + \theta_k \right) \right)}
\right] \, , 
\label{A2B1oversin2}
\end{align}
\begin{align}
  & \scalar{ \left(A_2' - \delta_2'\right) B_1'}{e_l} 
\nonumber\\
= & - \sum_{\scriptsize{\begin{matrix} i,j,k\\i \neq j \end{matrix}}} a_i a_j a_k H_{ijkl} 
\left[ 
\frac{\omega_j}{\omega_j - \omega_i} ( \cos \stackrel{i \leftrightarrow k}{\left( \theta_i - \theta_j - \theta_k \right)} + \cos \stackrel{j \leftrightarrow k}{\left( \theta_i - \theta_j + \theta_k \right)} ) \right.
\nonumber\\
& \hspace{33mm} \left. + \underline{\frac{\omega_j}{\omega_j + \omega_i} \left( \cos \left( \theta_i + \theta_j - \theta_k \right) + \cos\left( \theta_i + \theta_j + \theta_k \right) \right) }
\right]
\nonumber\\
 & -4 \sum_{i,j,k} a_i a_j a_k \tilde{X}_{klij} 
( \cos \stackrel{i \leftrightarrow k}{\left( \theta_i - \theta_j - \theta_k \right)} + \cos \stackrel{j \leftrightarrow k}{\left( \theta_i - \theta_j + \theta_k \right)} 
\nonumber\\
 & \hspace{33mm} \left. + \cos \left( \theta_i + \theta_j - \theta_k \right) + \cos\left( \theta_i + \theta_j + \theta_k \right) \right)
\nonumber\\
 & - \frac{1}{2} \sum_{i,k} a_i^2 a_k \left[ 
\left( 2 H_{iikl} + 4 \omega_i^2 M_{kli}\right) \cos(\theta_k) 
+ \underline{H_{iikl} \left( \cos \left( 2 \theta_i - \theta_k \right) + \cos\left( 2 \theta_i + \theta_k \right) \right)}
\right] \, , 
\label{primes}
\end{align}
\begin{align}
  & 2 \scalar{ A_2 \ddot B_1}{e_l} 
\nonumber\\
= & 2 \sum_{\scriptsize{\begin{matrix} i,j,k\\i \neq j \end{matrix}}} a_i a_j a_k \omega_k^2 X_{ijkl} 
\left[ 
\frac{\omega_j}{\omega_j - \omega_i} ( \cos \stackrel{i \leftrightarrow k}{\left( \theta_i - \theta_j - \theta_k \right)} + \cos \stackrel{j \leftrightarrow k}{\left( \theta_i - \theta_j + \theta_k \right)} ) \right.
\nonumber\\
& \hspace{33mm} \left. + \underline{\frac{\omega_j}{\omega_j + \omega_i} \left( \cos \left( \theta_i + \theta_j - \theta_k \right) + \cos\left( \theta_i + \theta_j + \theta_k \right) \right) }
\right]
\nonumber\\
 & + \sum_{i,k} a_i^2 a_k \omega_k^2 \left[ 
\left( 2 X_{iikl} + 4 \omega_i^2 W_{klii}\right) \cos(\theta_k) 
+ \underline{X_{iikl} \left( \cos \left( 2 \theta_i - \theta_k \right) + \cos\left( 2 \theta_i + \theta_k \right) \right)}
\right] \, , 
\label{A2ddotB1}
\end{align}
\begin{align}
  & \scalar{ \dot A_2 \dot B_1}{e_l} 
\nonumber\\
= & - \sum_{i,j,k} a_i a_j a_k \omega_j \omega_k X_{ijkl} 
\left( 
\cos \left( \theta_k - \theta_j - \theta_i \right) + \cos \stackrel{j \leftrightarrow k}{\left( \theta_k - \theta_j + \theta_i \right)} \right.
\nonumber\\
& \hspace{37mm} \left. - \cos \stackrel{i \leftrightarrow k}{\left( \theta_k + \theta_j - \theta_i \right)} - \cos \left( \theta_k + \theta_j + \theta_i \right) 
\right) \, ,
\label{dotA2dotB1}
\end{align}
\begin{align}
  & -2 \scalar{ \delta_2 \ddot B_1}{e_l} 
\nonumber\\
= & \sum_{\scriptsize{\begin{matrix} i,j,k\\k \neq l \end{matrix}}} a_i a_j a_k \frac{\omega_k^2}{\omega_l^2 - \omega_k^2}  
\left[ 
Z^+_{ijkl} ( \cos \stackrel{i \leftrightarrow k}{\left( \theta_i - \theta_j - \theta_k \right)} + \cos \stackrel{j \leftrightarrow k}{\left( \theta_i - \theta_j + \theta_k \right)} ) \right.
\nonumber\\
& \hspace{31mm} \left. - Z^-_{ijkl} \left( \cos \left( \theta_i + \theta_j - \theta_k \right) + \cos\left( \theta_i + \theta_j + \theta_k \right) \right)
\right]
\nonumber\\
 & - \sum_{i,j} a_i a_j a_l \omega_l^2 
\left[ 
\left( \omega_i \omega_j P_{ijl} + B_{ijl} \right) \left( \cos \left( \theta_i - \theta_j - \theta_l \right) + \cos \left( \theta_i - \theta_j + \theta_l \right) \right) \right. 
\nonumber\\
& \hspace{22mm} \left. - \left( \omega_i \omega_j P_{ijl} - B_{ijl} \right) \left( \cos \left( \theta_i + \theta_j - \theta_l \right) + \cos \left( \theta_i + \theta_j + \theta_l \right) \right) 
\right] \, , 
\label{delta2ddotB1}
\end{align}
\begin{align}
  & - \scalar{ \dot \delta_2 \dot B_1}{e_l} 
\nonumber\\
= & - \frac{1}{2} \sum_{\scriptsize{\begin{matrix} i,j,k\\k \neq l \end{matrix}}} a_i a_j a_k \frac{\omega_k}{\omega_l^2 - \omega_k^2}  
\left[ 
\left( \omega_i - \omega_j \right) Z^+_{ijkl} ( \cos \stackrel{i \leftrightarrow k}{\left( \theta_i - \theta_j - \theta_k \right)} - \cos \stackrel{j \leftrightarrow k}{\left( \theta_i - \theta_j + \theta_k \right)} ) \right.
\nonumber\\
& \hspace{38mm} \left. + \left( \omega_i + \omega_j \right) Z^-_{ijkl} \left( -\cos \left( \theta_i + \theta_j - \theta_k \right) + \cos\left( \theta_i + \theta_j + \theta_k \right) \right)
\right]
\nonumber\\
 & + \frac{1}{2} \sum_{i,j} a_i a_j a_l \omega_l 
\left[ 
\left( \omega_i - \omega_j \right) \left( \omega_i \omega_j P_{ijl} + B_{ijl} \right) \left( \cos \left( \theta_i - \theta_j - \theta_l \right) - \cos \left( \theta_i - \theta_j + \theta_l \right) \right) \right. 
\nonumber\\
& \hspace{24mm} \left. + \left( \omega_i + \omega_j \right) \left( \omega_i \omega_j P_{ijl} - B_{ijl} \right) \left( -\cos \left( \theta_i + \theta_j - \theta_l \right) + \cos \left( \theta_i + \theta_j + \theta_l \right) \right) 
\right] \, , 
\label{dotdelta2dotB1}
\end{align}
where the interaction coefficient $Z^{\pm}_{ijkl}$ is defined in (\ref{Zijkl}). The sum of the underlined terms in eqs. (\ref{A2B1oversin2}-\ref{A2ddotB1}) gives:
\begin{align}
& \sum_{\scriptsize{\begin{matrix} i,j,k\\i \neq j \end{matrix}}} \frac{\omega_j}{\omega_j + \omega_i} a_i a_j a_k \left( -8 \tilde{X}_{ijkl} - H_{ijkl} + 2 \omega_k^2 X_{ijkl} \right) 
\left( \cos \left( \theta_i + \theta_j - \theta_k \right) + \cos\left( \theta_i + \theta_j + \theta_k \right) \right) 
\nonumber\\
 & \hspace{10mm} + \frac{1}{2} \sum_{i,k} a_i^2 a_k \left( -8 \tilde{X}_{iikl} - H_{iikl} + 2 \omega_k^2 X_{iikl} \right) 
\left( \cos \left( 2 \theta_i - \theta_k \right) + \cos\left( 2 \theta_i + \theta_k \right) \right) 
\nonumber\\
 = & \sum_{i,j,k} \frac{\omega_j}{\omega_j + \omega_i} a_i a_j a_k \left( -8 \tilde{X}_{ijkl} - H_{ijkl} + 2 \omega_k^2 X_{ijkl} \right) 
\left( \cos \left( \theta_i + \theta_j - \theta_k \right) + \cos\left( \theta_i + \theta_j + \theta_k \right) \right) 
\end{align}
Now, interchanging indices in some terms as indicated in eqs. (\ref{A2B1oversin2}-\ref{dotdelta2dotB1}), we finally get (\ref{completeS3l}).
\newpage
One can prove that for $\scalar{{1\over \sin^2 x} B_1 B_2}{e_l}$ all secular $(+++)$ and $(+--)$ terms vanish. After simplifying we get 
\begin{align}
& -\int c_i(t') c_j(t') \sin {(\omega_k t') dt' |_{t'=t} \cos {(\omega_k t)}} +\int c_i(t') c_j(t') \cos {(\omega_k t') dt' |_{t'=t} \sin {(\omega_k t)}} \nonumber \\
&= {1\over 4} a_i a_j \left(-\frac{\cos(\theta_i-\theta_j)}{\omega_i-\omega_j-\omega_k}-\frac{\cos(\theta_i+\theta_j)}{\omega_i+\omega_j-\omega_k}+\frac{\cos(\theta_i-\theta_j)}{\omega_i-\omega_j+\omega_k}+\frac{\cos(\theta_i+\theta_j)}{\omega_i+\omega_j+\omega_k} \right) \; ,
\end{align}
where $\omega_i-\omega_j-\omega_k \neq 0$, $\omega_i+\omega_j-\omega_k \neq 0$, $\omega_i-\omega_j+\omega_k \neq 0$, $\omega_i+\omega_j+\omega_k \neq 0$. Multiplying by $c_m(t)/\omega_k = a_m \cos(\theta_m)/\omega_k$, we obtain
\begin{align}
&- {1\over 4} a_i a_j a_m \left(\frac{\cos(\theta_i-\theta_j+\theta_m)}{(\omega_i-\omega_j)^2-\omega_k^2}+\frac{\cos(\theta_i+\theta_j-\theta_m)}{(\omega_i+\omega_j)^2-\omega_k^2} \right. \nonumber \\
&\qquad \left. +\frac{\cos(\theta_i-\theta_j-\theta_m)}{(\omega_i-\omega_j)^2-\omega_k^2} + \frac{\cos(\theta_i+\theta_j+\theta_m)}{(\omega_i+\omega_j)^2-\omega_k^2} \right)\; .
\end{align}
This expression is multiplied by $K_{ijk} K_{kml}$. Using~\eqref{K0condition} we get:

1) for the $(+++)$ terms:
\begin{equation*}\omega_i + \omega_j + \omega_m = \omega_l \Rightarrow i+j+m+6 = l \;,
\end{equation*}
\begin{equation*}
k > i+j+2 \Rightarrow K_{ijk}=0 \;,
\end{equation*}
\begin{equation*}
l > k+m+2 \Rightarrow k < i+j+4 \Rightarrow K_{kml}=0 \;,
\end{equation*}
which means that for every $k \in \mathbb{N}$ at least one of the conditions $K_{ijk}=0$ or $K_{kml}=0$ is satisfied, so the whole term always vanishes. 

2) for the $(+--)$ terms:
\begin{equation*}\omega_i - \omega_j - \omega_m = \omega_l \Rightarrow i-j-m-6 = l \;,
\end{equation*}
\begin{equation*}
i > j+k+2 \Rightarrow k<i-j-2 \Rightarrow K_{ijk}=0 \;,
\end{equation*}
\begin{equation*}
k>l+m+2 \Rightarrow k>i-j-4 \Rightarrow K_{kml}=0 \;,
\end{equation*}
or
\begin{equation*}\omega_i + \omega_j - \omega_m = - \omega_l \Rightarrow i+j-m+6 = -l \;,
\end{equation*}
\begin{equation*}
k>i+j+2 \Rightarrow K_{ijk}=0 \;,
\end{equation*}
\begin{equation*}
m>l+k+2 \Rightarrow k<i+j+4 \Rightarrow K_{kml}=0 \;,
\end{equation*}
so the whole term always vanishes as in the previous case. A similar analysis for the $(++-)$ case does not imply that $K_{ijk}K_{kml}$ always equals 0. 
For the special cases $\omega_i-\omega_j-\omega_k = 0$, $\omega_i+\omega_j-\omega_k = 0$, $\omega_i-\omega_j+\omega_k = 0$ the secular terms vanish as $K_{ijk} = 0$. There are no terms in the sum that satisfy $\omega_i+\omega_j+\omega_k = 0$.

For the lower limit $t' = 0$ secular terms would appear if $\omega_k \pm \omega_m =  \pm \omega_l$. In this case $K_{kml} = 0$, so there is no contribution.

There are also no secular terms in $\langle {1\over \sin^2 x} B_1^3,e_l\rangle$ for the $(+++)$ and the $(+--)$ case. Equations obtained in the way analogous to~\eqref{Kjkn}--\eqref{K0condition}:
\begin{equation} \label{Gjkn}
G_{ijkn} = \int^{\pi/2}_0 e_i(x) e_j(x) e_k(x) e_n(x) \frac{\sin x}{\cos^3 x} \; ,
\end{equation}
\begin{equation}
G_{ijkn} \sim \int^1_{-1} (1-y)(1+y)^4 P_i^{(3,2)} P_j^{(3,2)} P_k^{(3,2)}\frac{d^n}{d x^n}\left( (1-y)^{3+n}(1-y)^{2+n} \right)
\end{equation}
imply that $G_{ijkn} = 0$ if 
\begin{equation} 
n>i+j+k+5
\end{equation} (because $(1-y)(1+y)^4 P_i^{(3,2)} P_j^{(3,2)} P_k^{(3,2)}$ is the polynomial of  order $i+j+k+5$). For the resonant terms $\omega_n = \omega_j+\omega_k+\omega_i$, hence $n=6+i+j+k$, which means that the resonant $(+++)$ or $(+--)$ terms in $\langle {1\over \sin^2 x} B_1^3,e_l\rangle$ always vanish.

We read the coefficients $Q_{ijkl}$, $U_{ijkl}$, $S_{ijkl}$, $R_{il}$, $T_l$ in (\ref{S3lfinal}) from the source term~\eqref{completeS3l} and apply identities (\ref{HMidentities})
to finally get simplified expressions~\eqref{Qijkl},~\eqref{Uijkl},~\eqref{Sijkl}, \eqref{Ril}, \eqref{Tl}.

\end{document}